\documentclass[a4paper,twocolumn,superscriptaddress,prl, aps, floatfix ]{revtex4-1}

\usepackage{epsfig}
\usepackage{verbatim}	 
\usepackage{amsmath,amssymb}
\usepackage{calrsfs}
\usepackage{times}
\usepackage{amsmath}
\usepackage{psfrag}
\usepackage{color}
\usepackage[latin1]{inputenc}
\usepackage[english]{babel}
\usepackage{graphicx}
\usepackage{braket}
\usepackage{natbib}
\usepackage{hyperref}
\usepackage{lipsum}
\usepackage{multirow}
\usepackage{cleveref}
\usepackage[colorinlistoftodos]{todonotes}

\addto\captionsenglish{}

\DeclareMathAlphabet{\mathcal}{OMS}{cmsy}{m}{n}

\begin{document}
\title{Prediction of scientific collaborations through multiplex interaction networks}

\author{Marta Tuninetti}
\affiliation{ISI  Foundation,  via  Chisola  5,  10126  Torino, Italy}

\author{Alberto Aleta}
\affiliation{ISI  Foundation,  via  Chisola  5,  10126  Torino, Italy}

\author{Daniela Paolotti}
\affiliation{ISI  Foundation,  via  Chisola  5,  10126  Torino, Italy}

\author{Yamir Moreno}
\affiliation{Institute for Biocomputation and Physics of Complex Systems (BIFI), University of Zaragoza, 50018 Zaragoza, Spain}
\affiliation{Department of Theoretical Physics, University of Zaragoza, 50018 Zaragoza, Spain}
\affiliation{ISI  Foundation,  via  Chisola  5,  10126  Torino, Italy}

\author{Michele Starnini}
\thanks{Corresponding author: michele.starnini@gmail.com}
\affiliation{ISI  Foundation,  via  Chisola  5,  10126  Torino, Italy}

\begin{abstract}

Link prediction algorithms can help to understand the structure and dynamics of scientific collaborations and  the evolution of Science. However, available algorithms based on similarity between nodes of collaboration networks are bounded by the limited amount of links present in these networks. 
In this work, we reduce the latter intrinsic limitation by generalizing the Adamic-Adar method to multiplex networks composed by an arbitrary number of layers, that encode diverse forms of scientific interactions. 
We show that the new metric outperforms other single-layered, similarity-based scores and that scientific credit, represented by citations, and common interests, measured by the usage of common keywords, can be predictive of new collaborations.
Our work paves the way for a deeper understanding of the dynamics driving scientific collaborations, and provides a new algorithm for link prediction in multiplex networks that can be applied to a plethora of systems. 

\end{abstract}

\maketitle

One of the main drivers of scientific discoveries is the establishment of new collaborations among researchers. 
The collision of two scientific trajectories, even if they belong to the same research area, brings in contact different methods, concepts, and ideas, creating the ideal environment for scientific production. 
Understanding the dynamics that drives the development of scientific collaborations is thus pivotal to
characterize the structure and evolution of science \cite{fortunato2018science}. In this endevour, two factors play a crucial role.  
On the one hand, the digitalization of large scale bibliographic databases has provided comprehensive data sets of publication records covering research from all disciplines, without geographical limits. 
By leveraging on these databases \cite{clauset2017data}, researchers have pictured the structure of different research fields \cite{sinatra2015century, battiston2019taking}, measured the emergence of new interdisciplinary areas \cite{53e9b58db7602d97040cf45d, RePEc:eee:infome:v:5:y:2011:i:1:p:87-100}, mapped the evolution of scientific interests \cite{doi:10.1177/0003122415601618, Aleta:2019aa}, and characterized scientific productivity at the individual and geographical level \cite{Bornmann2011, zhang2013characterizing, deville2014career, sinatra2016quantifying}. 
On the other hand, network science \cite{newman10-1} has been established as the main tool to analyze and model cooperation in science.  
Since the seminal work by Newman \cite{Newman2001Jan}, scientific collaborations are represented in the form of a network, where nodes stand for scientists and a link between two nodes is drawn if two scientists co-authored a paper together. 

The forecast of a new collaboration translates, within the network science domain, into a link prediction problem \cite{Nowell2003}, a prolific area of network research with applications ranging from detecting hidden links in economic networks \cite{Anand2018Apr} to enhance the user experience in online social platforms \cite{Wang2015Jan}.
Link prediction algorithms can be classified mainly into two categories: similarity based methods and probabilistic models \cite{Marjan2018Oct,MartnnezVnctor2016Dec}. 
Since the latter can be computationally unfeasible for large networks, a lot of attention has been devoted to the creation of good similarity scores. Many of these similarity methods are based on the same basic idea, two nodes are likely to be linked if they share a common neighbor \cite{Bliss2014Sep, Adamic2003Jul}. 
Despite its simplicity, this concept has proven to be quite useful for highly assortative networks, such as scientific collaboration networks \cite{Clauset2008May}. Moving from this simple approach, several attempts have been made to improve the prediction of collaborations by incorporating more data. For instance, by adding information about the organization the authors work at \cite{Cho2018Dec}, topical interest \cite{Hassan2006}, time at which collaborations are established \cite{Moradabadi2017Sep}, offline relationships among employees of the same university \cite{Najari2019Dec}, weights of the collaboration links \cite{Sett2016Jan} or journal information \cite{ZhangJinzhu2017Jan}. 
However, in most of these approaches, the scores are computed individually for each set of data and then aggregated into a unique score, possibly after associating a specific weight to each set. 
Furthermore, as signaled by Jia et al. \cite{Jia2020}, the prediction power of any similarity-based link prediction algorithm is bounded due to the limited amount of links present in the network. 

In this paper, we reduce this intrinsic limitation and take advantage of the whole information included in publication records to improve the prediction of new collaborations. To do so, we represent \emph{scientific interaction networks} as a multiplex network \cite{Kivela2014Sep,Aleta2019Mar}, where nodes represent scientists and different kinds of interactions among them are encoded in different layers. 
We propose a novel metric for link prediction in multiplex networks, based on a generalization of the Adamic-Adar method for single-layered networks \cite{Adamic2003Jul}, which leverages on the multiplex representation of scientific interaction networks. 
While our method is  general and can be applied to any multiplex topology, composed by an arbitrary number of layers, here we focus on scientific credit, represented by citations, and common interests, measured by the usage of common keywords, to predict new collaborations. 
Our metric fully exploits the complexity of the relationships that might be established across the different scientific interactions, by considering all possible triadic closures in the corresponding multiplex representation. We show that this score is able to outperform other scores based on single-layered similarity between nodes, demonstrating the prediction capabilities of scientific credit and  common interests in the development of future  collaborations. 

Our data set is composed by merging two different bibliographical  sources: the  American Physical Society (APS) database, including the authors, publication date and references of over 400,000 papers published from 1893 to 2009 \cite{Radicchi2009Nov}, and the ArnetMiner database \cite{OAG1}, containing title, author list, publication year and keywords for almost 155 billion papers belonging to multiple research fields. From this data set, we select only those papers present in the APS data set, by matching the DOI number. 
Our final data set is composed, for each paper, by the list of authors with their affiliations, the list of keywords associated to the paper, and the papers cited as references. 
Before analyzing the data, we apply a cleaning procedure to the keywords obtained,  see Supplementary Material (SM) for details. 

We then build a \text{scientific interaction network}, in the form of a weighted multiplex network \cite{10.1371/journal.pone.0097857}, where nodes represent scientists, and different layers account for different interactions among them: collaboration, common interest, and scientific credit.
The first layer ($c$) represents collaboration, and it corresponds to classical co-authorship networks: two authors are linked if they published at least a paper together. The second layer ($r$) represents scientific credit, measured by references or citations: a link between two authors $u$ and $v$ indicates that $u$ cited at least one paper from $v$. Even though this network is in principle directed, for our purpose we will treat it as undirected. Lastly, the third layer ($k$) represents common scientific interests, which can be measured by the usage of common keywords: two authors are connected if, out of all the keywords they have ever used, they have at least one in common. 
Therefore, the weight $w_{uv}^{\alpha}$ represents the number of co-authored papers ($\alpha=c$), citations ($\alpha=r$), or common keywords ($\alpha=k$) between two authors $u$ and $v$.

We consider two subsequent time intervals, the first over which link prediction algorithms will be trained, corresponding to a training network with all authors who published a paper between $t_0$ and $t_1$, and a test interval for testing the predictions of new collaborations, including all authors active between $t_1$ and $t_2$. We then extract the \emph{Core} of these networks, corresponding to the authors that have at least $k_{min}$ edges both in the training and test intervals, to ensure authors to be active in both intervals, as standard in link prediction problems on social networks \cite{Nowell2003}.
In order to reduce the computational complexity of the prediction algorithms, we restrict our analysis to only papers published in Physical Review Letters (PRL) between $t_0= 1994$ and $t_2=2005$, split at  $t_1=2000$, see SM for details on how we choose the intervals. 
The resulting scientific interaction network is composed by $N=24,366$ authors, further details  are reported in the SM. 
Here, we set $k_{min}=3$, so the number of new links to be predicted is equal to $E_p=7,563$. 
In the SM, we show results for link prediction with a   \emph{Core} obtained by setting $k_{min}=5$.

The most common and successful method for link prediction in social networks has been developed by Adamic and Adar ($AA$) \cite{Adamic2003Jul}. 
The $AA$ score between nodes $u$ and $v$ is given by the number of common neighbors weighted by their degree, 
\begin{equation}\label{eq:AA}
AA(u,v) = \sum_{w\in \Gamma(u) \cap \Gamma(v)} \frac{1}{\ln (k_w) }\,.
\end{equation}
where $\Gamma(u)$ represents the set of neighbors of node $u$ and $k_w = |\Gamma(w)|$ is the degree of node $w$.
Note that the common neighbors of $u$ and $v$ can be both in the \emph{Core} and outside it.
In a multilayer network, the $AA$ score could be applied to different layers, depending on which layer $\alpha$ the set of neighbors $w \in \Gamma_{\alpha}(u) \cap \Gamma_{\alpha}(v)$ is considered, where $\Gamma_{\alpha}(u)$ represents the set of neighbors of node $u$ in layer $\alpha$.
For the sake of comparison, we first apply the $AA$ score to all layers, separately, as single-layered scores to predict links in the collaboration layer. 
The rationale is that two scientists are more likely to collaborate if they share many common collaborators (AA in the collaboration layer, or $AA_c$), cite the same set of authors ($AA_r$ in the reference layer), or have similar scientific interests ($AA_k$ in the keyword layer). Note, however, that this does not fully exploits the richness of a multilayer representation.

The quality of the prediction of different scores can be quantified by two metrics: the Receiver Operating Characteristics (ROC) curve, with the corresponding Area Under the Curve (AUC) value, and the Precision. The Precision can be computed as $n^\ast/n$, where $n$ is the number of new links that we want to predict and $n^\ast$ is the amount of correct predictions among the top $n$ links.  Thus, it provides complementary information to the one given by the AUC.
It is important to highlight that, due to the limited amount of links present in a network, the AUC of any similarity-based link prediction algorithm is bounded \cite{Jia2020}. For instance, if similarity is based on common neighbors, two nodes without any neighbor in common will have a score equal to zero. The number of scoreless links bounds the maximum and minimum values of the AUC to $\text{AUC}_\text{min} = \frac{1}{2}(1+p_1)(1-p_2)$ and $\text{AUC}_\text{max} = \text{AUC}_\text{min} + p_1 p_2$, where $p_1$ ($p_2$) is the fraction of links with a score different from 0 among those links that will (will not) exist in the future, see SM for details. Note that only when $p_1 = p_2 = 1$, i.e., there are no scoreless links, it holds $\text{AUC}_\text{min} = 0$ and $\text{AUC}_\text{max} =1$.

In Table \ref{tab:AUC_multiplex}, we show the Precision and AUC values obtained for the AA method applied to each layer, together with the theoretical bounds of the AUC. Interestingly, the  $AA_c$ score has an AUC value quite close to the random one, but the highest Precision among single-layered scores. This reflects the fact that, even though the heuristics behind the metric seem to be a good proxy of the real dynamics, the limited amount of information hinders the prediction process. On the other hand, the keywords layer is the more dense one and thus  it carries much more information than the others, yielding a larger theoretical maximum for the AUC of the $AA_k$ score. However, its Precision is not as good as other metrics, indicating that sharing keywords is not such a good descriptor of the dynamics behind establishing new collaborations. The $AA_r$ method shows a behavior between the other two: the citation layer carries less information than the keyword layer but more than the collaboration one, as shown by the larger AUC value of the $AA_r$ method with respect to the $AA_c$ one. We also considered other single-layered scores built on the citation and keyword layers, such as mutual citations between two scientists, directed citations normalized by the total citations received and given by each scientist, and number of common keywords normalized by the keywords frequency. In the SM we show the ROC curves, the AUC and the Precision of these metrics, demonstrating that they cannot outperform the simpler metrics showed here. 

\begin{table}[]
\begin{tabular}{lccc}
\hline
\textbf{Method}	& \textbf{Precision} & \textbf{AUC} & \textbf{AUC [worst-best]} \\
\hline
Random   & 4.3$\cdot10^{-4}$ & 0.5 	& [0.50-0.50] \\
$AA_c$   	& 0.041                  & 0.5635          & [0.5633-0.5636]                 \\
$AA_r$   	& 0.017                  & 0.6361          & [0.6282-0.6393]                 \\
$AA_k$   & 0.006                  & 0.6481           & [0.0171-0.9951]               \\
$MAA$ (all triads)      & 0.042                  & 0.7620           & [0.0147-0.9971]              \\
Aggregated               	& 0.006                  & 0.6495           & [0.0147-0.9971]              \\
\hline \\
\end{tabular}
\caption{Precision and AUC values obtained for different metrics proposed, with the theoretical bounds of the AUC. 
We consider each layer separately and the $MAA$ score given by Eq. \eqref{eq:MAA}, with coefficients $\eta_{ck}=0.05$ and $\eta_{cr}=0.1$ which maximize both AUC and Precision (see Fig. \ref{fig:phsp}).
Finally, we compare with a aggregated, single-layered network given by the projection of all layers onto a single layer. Note that in this case we obtain almost the same values of the $AA_k$ score.}
\label{tab:AUC_multiplex}
\end{table}

\begin{figure}
\includegraphics[width=\linewidth]{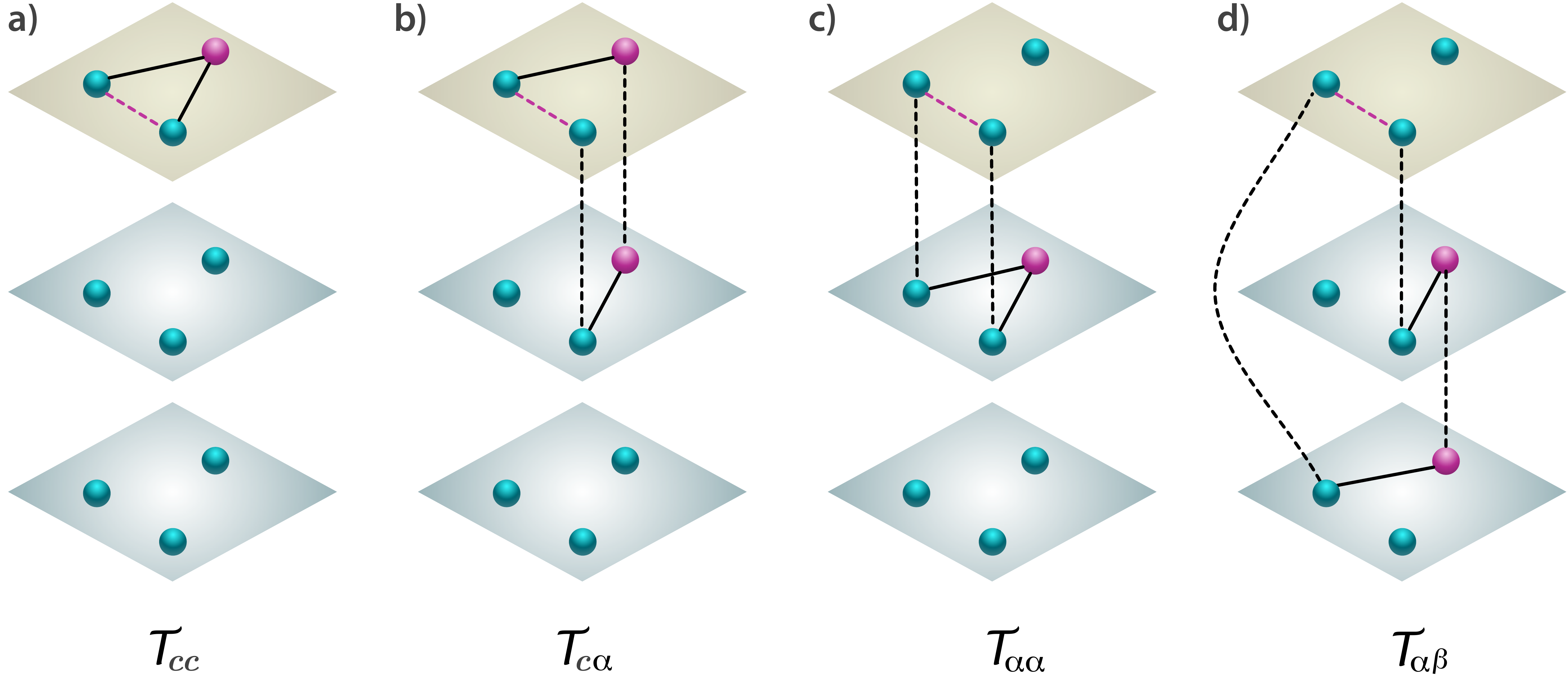}
\caption{\textbf{Triadic relationships in a multiplex network.} Given two nodes $u$ and $v$ for which we want to predict the future existence of a link (red dashed line) in the top layer (green), based on their connections with another node $w$ (pink) via triadic closure, we can distinguish four types of triadic relationships: (a) $u$ and $v$ are both connected to $w$ in the prediction layer; (b) the link between $u$ and $w$ is in the prediction layer, but $v$ and $w$ are connected in a different layer, or viceversa; (c) both $u$ and $v$ are connected to $w$ in a layer different from the prediction layer; (d) $u$ and $w$ are connected in a layer different from the one in which we want to make our prediction and $v$ and $w$ are connected in a third layer different from those two.}
\label{fig:trian}
\end{figure}

The differences in the AUC and Precision of different layers show the need to go beyond single-layered scores and combine them into a more general metric, that fully exploits the multiplex nature of the scientific interaction networks. Note, indeed, that single-layered metrics considered triadic relations among three nodes $u$, $v$ and $w$, in which the link $u-v$ to be predicted lays in the collaboration layer, while two links $u-w$ and $v-w$ lay both in the same layer, such as in the case of the $AA_r$ score, where the two links lay in the citation layer.  However, triadic relations in multiplex networks can be far more richer \cite{PhysRevE.89.032804, Cozzo2015Jul}. Figure \ref{fig:trian} shows different kinds of triadic relations in multiplex networks, in which, for the sake of clarity, we assumed $\alpha \neq \beta \neq c$. Given that the link $u-v$  to be predicted must be in the collaboration layer, one can distinguish four types of triadic relations depending on the location of the $(u,w)$ and $(v,w)$ links: 
i) $\mathcal{T}_{cc} = \{(u,v,w) | w \in \Gamma_{c}(u) \cap \Gamma_{c}(v)  \}$, in which both links lay in the collaboration layer $c$; 
ii) $\mathcal{T}_{c\alpha} = \{(u,v,w) | w \in \Gamma_{c}(u) \cap \Gamma_{\alpha}(v) \}$ and $\mathcal{T}_{\alpha c}$, in which one lays in the collaboration layer and the other is either in the citation $\alpha=r$ or keywords $\alpha=k$ layer; 
iii) $\mathcal{T}_{\alpha \alpha} =  \{(u,v,w) | w \in \Gamma_{\alpha}(u) \cap \Gamma_{\alpha}(v)  \}$, in which both links are in the same layer $\alpha$ different from the collaboration one; 
iv) $\mathcal{T}_{\alpha\beta} =  \{(u,v,w) | w \in \Gamma_{\alpha}(u) \cap \Gamma_{\beta}(v) \} $ and $\mathcal{T}_{\beta\alpha}$, each link lays in a different layer, different from the collaboration one. 

Within this formalism, one can consider a score that counts the common neighbors closing triads of each type, and weight each contribution by the logarithm of the degree, as in the Adamic-Adar score,
\begin{equation}\label{eq:MAA}
MAA(u,v)  =  \sum_{\alpha, \beta}    \sum_{w\in\mathcal{T}_{\alpha\beta}}  \frac{\eta_{c\alpha} \eta_{c\beta}}{\sqrt{\langle k \rangle_\alpha \langle k \rangle_\beta}} 		 \frac{1}{\sqrt{\ln(k_w^\alpha ) \ln(k_w^\beta)}}
\end{equation}
This expression is the generalization of the Adamic-Adar score for multiplex networks (MAA) with an arbitrary number of layers, in which the links to be predicted all lay in the same layer $c$. For our purpose, we assume three layers, so $\alpha, \beta \in \{c,r,k\}$. Several considerations are in order. First, the original AA score in collaboration networks (case (a) of Fig. \ref{fig:trian}) is recovered by considering only one layer, $\alpha= \beta = c$. The application of AA to other layers, indicated previously as $AA_r$ and $AA_k$, corresponds to triads closures  $\mathcal{T}_{\alpha\alpha}$ (case (c) of Fig. \ref{fig:trian}). Second, the contribution of each triads $(u,v,w) \in \mathcal{T}_{\alpha\beta}$ is weighted by the square root of the logarithm of the degree of node $w$ in the two layers involving $\alpha$ and $\beta$. With this choice, the original weight $1/ \ln(k_w)$ is naturally recovered for $\alpha=\beta=c$. Third, note that different layers of a multiplex network may show very different densities, as in the case of scientific interaction networks (see SM). In case of similarity scores based on the number of common neighbors, as in this case, denser layers will have more triads and thus will be less informative. We take into account this by weighting the contribution of each type of triadic relation by the square root of the average degree of the layers involved, $\sqrt{\langle k \rangle_\alpha}$. Fourth, the coefficients $\eta_{c\alpha}$ before each term allow us to control the relative weight of each type of triadic closure in the total score of the link. We choose them in a way that $\eta_{c\alpha}$ corresponds to the weight of layer $\alpha$. Without lack of generality, we choose $\sum_{\alpha} \eta_{c\alpha} = \eta_{cc} +  \eta_{cr} +  \eta_{ck}  = 1$. The case $\eta_{cc}=1$, $\eta_{cr} = \eta_{cr}  = 0$, corresponds to the original AA score on collaboration networks.

\begin{figure}
\includegraphics[width=\linewidth]{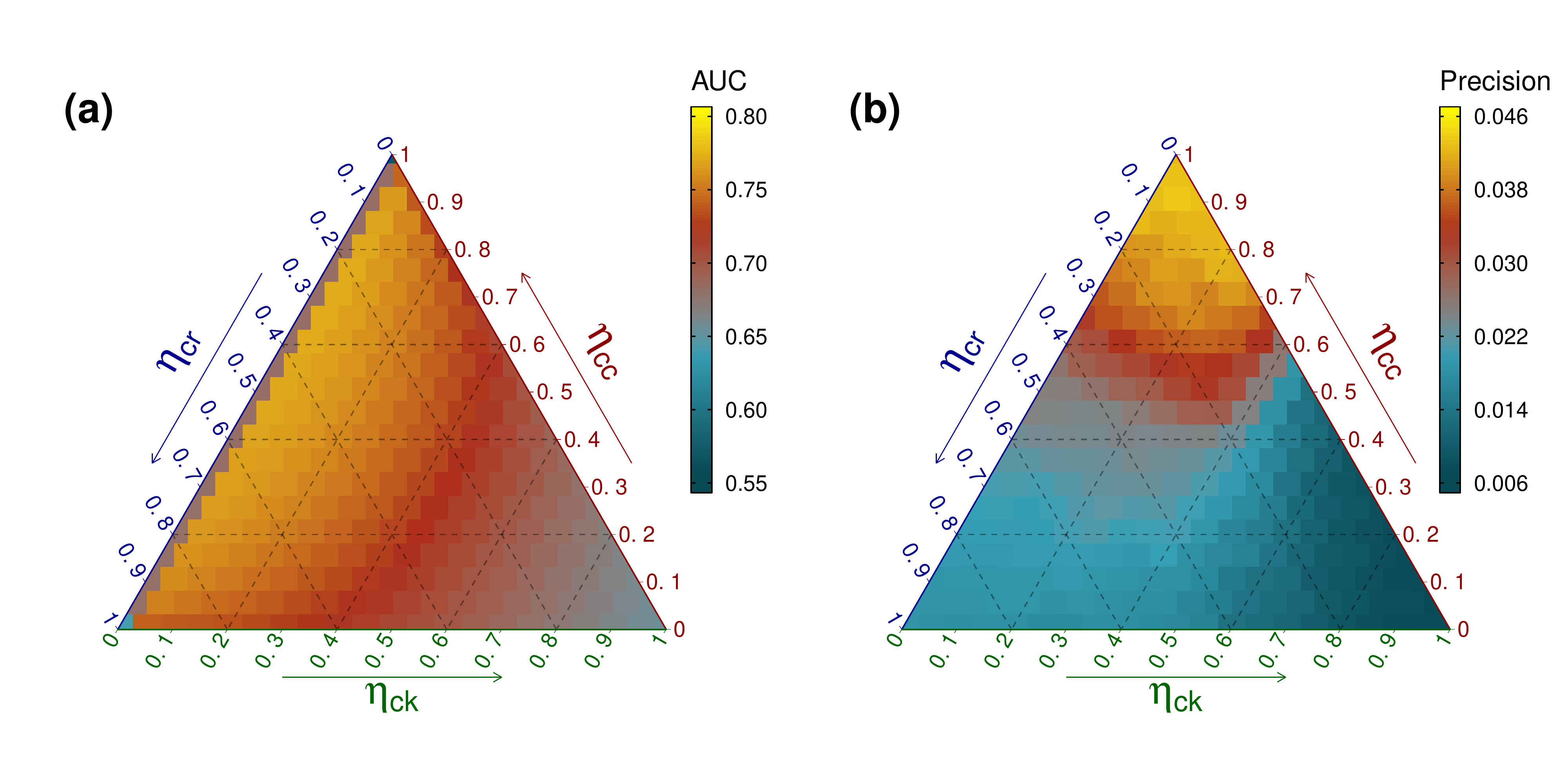}
\caption{\textbf{AUC and Precision values of the $MAA$ metric for different values of the coefficients $\eta_{c\alpha}$.} Varying the values of $\eta_{cr}$ and $\eta_{ck}$, the third parameter $\eta_{cc}$ is naturally fixed.}
\label{fig:phsp}
\end{figure}

Figure \ref{fig:phsp} shows the AUC and Precision of the $MAA$ metric, given by Equation \eqref{eq:MAA}, as a function of the coefficients $\eta_{c\alpha}$.  
Figure \ref{fig:phsp}(a) shows that the AUC value receives an important contribution from triads involving the citations and keywords layers, as shown by the discontinuity for $\eta_{cc}<1$. This result is consistent with the fact that citations and keywords relationships contribute to increase the amount of information carried by the co-authorship layer, see Table \ref{tab:AUC_multiplex}. The Precision is maximum for  $\eta_{ck}=0.05$ and $\eta_{cr}=0.1$ (see Figure \ref{fig:phsp}(b)), showing that the contribution of the collaboration layer is important to keep high precision. 

\begin{figure}
\includegraphics[width=\linewidth]{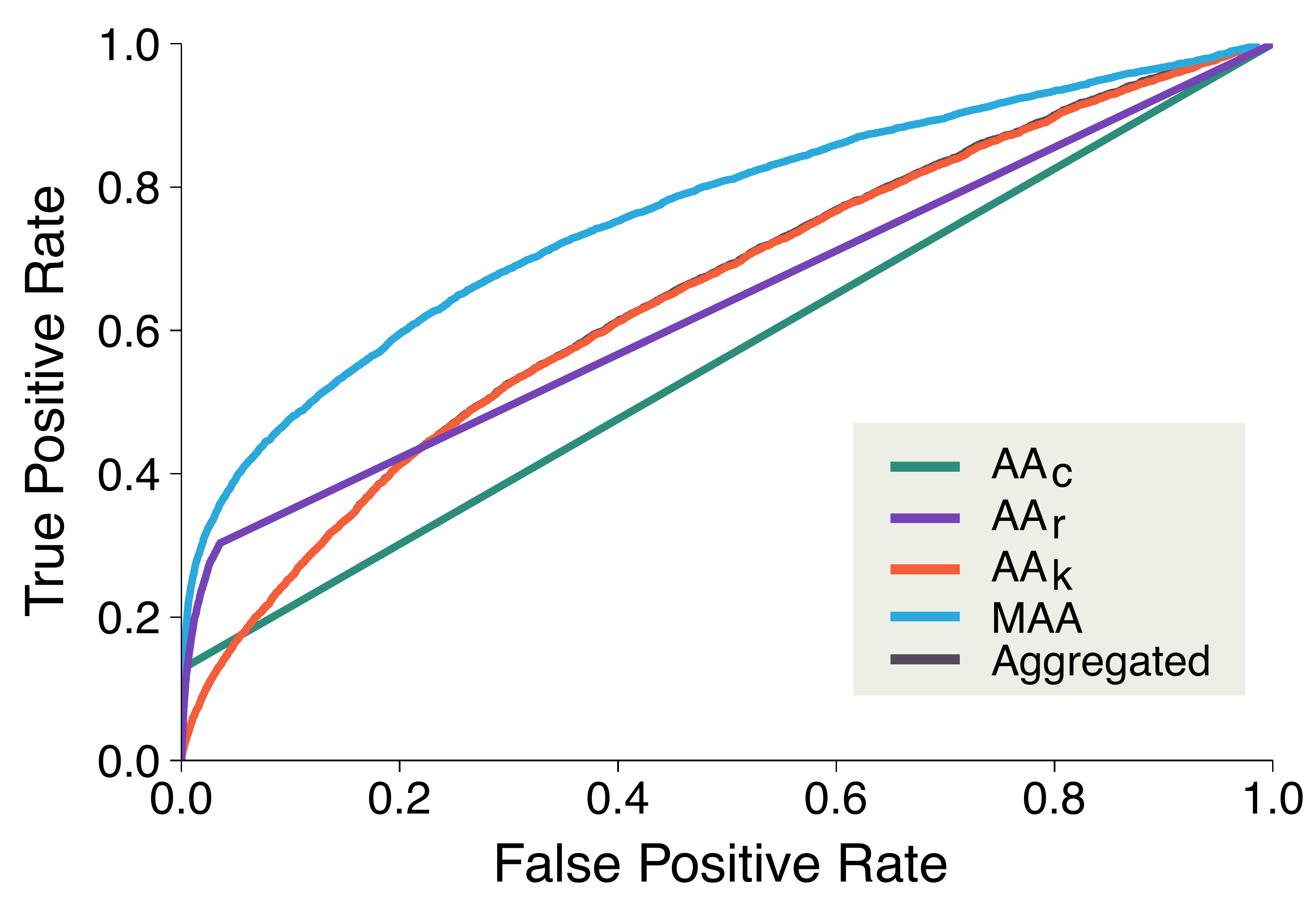}
\caption{\textbf{ROC curves obtained for single layered scores, for the $MAA$ metric, and for the aggregated network.} 
The $MAA$ metric with coefficients $\eta_{ck}=0.05$ and $\eta_{cr}=0.1$ shows the best perfomance. 
The aggregated network curve is not visible, covered by the $AA_k$ curve, showing that the main contribution in the aggregated network comes from the keywords layer.}
\label{fig:ROC_multiplex}
\end{figure}

In Table \ref{tab:AUC_multiplex} we show that the $MAA$ metric given by Equation \eqref{eq:MAA} with coefficients $\eta_{ck}=0.05$ and $\eta_{cr}=0.1$, which maximize both AUC and Precision, has a much larger  AUC and Precision than all other single layered metrics. We also compare with an aggregated, single-layered network given by the projection of all layers onto a single layer, showing that in this case we obtain almost the same values of the $AA_k$ score, given that the projected network is dominated by the keywords layer. Finally,  in Figure \ref{fig:ROC_multiplex} we compare the ROC curves obtained with each of the proposed metrics. These curves clearly show that, at some point of the ranking, for $AA_c$ and $AA_r$ metrics only scoreless links remain and therefore the curve follows a linear trend. The ROC curve of the aggregated networks is equivalent to the ROC curve of the $AA_k$ score, while the ROC curve  of the $MAA$ metric given by Eq. \eqref{eq:MAA} shows the best performance.

Before concluding, we stress that the metric encoded in Eq. \eqref{eq:MAA} is different from previous extensions of link prediction to multilayer networks. 
The idea of predicting links in multilayer networks has been explored during the last decade from several different points of view. For instance, Davis et al. \cite{Davis2013Jun} proposed a similar technique to include multi-relational data for link prediction, but from a probabilistic point of view. Similarly, several extensions of probabilistic models to multilayer networks have been proposed \cite{Kleineberg2016Jul,Hassan2006,Lu2010Dec,Matsuno2018,Pujari2015}. Other works, rather than focusing on incorporating new data to already existing networks, used multilayer structures to focus on the temporal evolution of the networks \cite{Hajibagheri2016Aug,Yao2016Jan}. Along the lines of our proposal, several works extended the notion of neighborhood to multilayer networks \cite{Jalili2016Nov,Hristova2016Dec,Mandal2018Nov,Junuthula2018}. However, they focused on networks of two layers without considering the triadic structures present in multilayer networks of at least 3 layers. Similarly, other approaches calculate the score of each layer and aggregate all of them (possibly with some weights), effectively neglecting structures of types $\mathcal{T}_{c\alpha}$ and $\mathcal{T}_{\alpha\beta}$ \cite{Sharma2015Nov,Yao2017Jul,Samei2019Oct}. It is also worth remarking that the metric given in  Eq. \eqref{eq:MAA} is general for an arbitrary number of layers.

To sum up, we showed that scientific credit and common scientific interests can be predictive of new collaborations between scientists. For this purpose, we reconstructed a data set of publication records by merging different bibliographic sources, including keywords that indicate the topics of papers. We represent this data set as a scientific interaction multiplex network, in which each layer encodes a different kind of interaction, directed or undirected. First, we show that single-layered link prediction metrics are bounded by the amount of information available, and that different layers can have high precision and low AUC, or viceversa. Then, we proposed a more general method that fully exploits the multiplex nature of the scientific interaction networks. Our metric is a generalization of the Adamic-Adar score for multiplex networks with an arbitrary number of layers, and it is able to outperform single-layered scores.

Our method permits to single out the contributions of scientific credit and common scientific interests in improving the prediction of new collaborations in different areas of Physics, and in particular with respect  to interdisciplinary collaborations. 
In future works, it would be interesting to incorporate additional information from publication records into the scientific interaction multiplex network, such as institutional affiliations and geographical locations, to see if these features are predictive of new collaborations. 
Our methodology could also be applied beyond the field of Physics, to understand if citations and keywords usage improve link prediction across scientific fields. 
On a different direction, it would be interesting to apply the $MMA$ metric for link prediction in other systems represented by multiplex networks, such as social networks in which each layer accounts for a different social interaction \cite{Starnini:2017aa}.

\begin{acknowledgments}
We acknowledge support from Intesa Sanpaolo Innovation Center. Y. M. acknowledges partial support from the Government of Arag\'on and FEDER funds, Spain through grant E36-17R to FENOL, and by MINECO and FEDER funds (grant FIS2017-87519-P). The funders had no role in study design, data collection, and analysis, decision to publish, or preparation of the manuscript.
\end{acknowledgments}

\bibliographystyle{apsrev4-1}

\begin{thebibliography}{54}%
\makeatletter
\providecommand \@ifxundefined [1]{%
 \@ifx{#1\undefined}
}%
\providecommand \@ifnum [1]{%
 \ifnum #1\expandafter \@firstoftwo
 \else \expandafter \@secondoftwo
 \fi
}%
\providecommand \@ifx [1]{%
 \ifx #1\expandafter \@firstoftwo
 \else \expandafter \@secondoftwo
 \fi
}%
\providecommand \natexlab [1]{#1}%
\providecommand \enquote  [1]{``#1''}%
\providecommand \bibnamefont  [1]{#1}%
\providecommand \bibfnamefont [1]{#1}%
\providecommand \citenamefont [1]{#1}%
\providecommand \href@noop [0]{\@secondoftwo}%
\providecommand \href [0]{\begingroup \@sanitize@url \@href}%
\providecommand \@href[1]{\@@startlink{#1}\@@href}%
\providecommand \@@href[1]{\endgroup#1\@@endlink}%
\providecommand \@sanitize@url [0]{\catcode `\\12\catcode `\$12\catcode
  `\&12\catcode `\#12\catcode `\^12\catcode `\_12\catcode `\%12\relax}%
\providecommand \@@startlink[1]{}%
\providecommand \@@endlink[0]{}%
\providecommand \url  [0]{\begingroup\@sanitize@url \@url }%
\providecommand \@url [1]{\endgroup\@href {#1}{\urlprefix }}%
\providecommand \urlprefix  [0]{URL }%
\providecommand \Eprint [0]{\href }%
\providecommand \doibase [0]{http://dx.doi.org/}%
\providecommand \selectlanguage [0]{\@gobble}%
\providecommand \bibinfo  [0]{\@secondoftwo}%
\providecommand \bibfield  [0]{\@secondoftwo}%
\providecommand \translation [1]{[#1]}%
\providecommand \BibitemOpen [0]{}%
\providecommand \bibitemStop [0]{}%
\providecommand \bibitemNoStop [0]{.\EOS\space}%
\providecommand \EOS [0]{\spacefactor3000\relax}%
\providecommand \BibitemShut  [1]{\csname bibitem#1\endcsname}%
\let\auto@bib@innerbib\@empty
\bibitem [{\citenamefont {Fortunato}\ \emph {et~al.}(2018)\citenamefont
  {Fortunato}, \citenamefont {Bergstrom}, \citenamefont {B{\"o}rner},
  \citenamefont {Evans}, \citenamefont {Helbing}, \citenamefont
  {Milojevi{\'c}}, \citenamefont {Petersen}, \citenamefont {Radicchi},
  \citenamefont {Sinatra}, \citenamefont {Uzzi} \emph
  {et~al.}}]{fortunato2018science}%
  \BibitemOpen
  \bibfield  {author} {\bibinfo {author} {\bibfnamefont {S.}~\bibnamefont
  {Fortunato}}, \bibinfo {author} {\bibfnamefont {C.~T.}\ \bibnamefont
  {Bergstrom}}, \bibinfo {author} {\bibfnamefont {K.}~\bibnamefont
  {B{\"o}rner}}, \bibinfo {author} {\bibfnamefont {J.~A.}\ \bibnamefont
  {Evans}}, \bibinfo {author} {\bibfnamefont {D.}~\bibnamefont {Helbing}},
  \bibinfo {author} {\bibfnamefont {S.}~\bibnamefont {Milojevi{\'c}}}, \bibinfo
  {author} {\bibfnamefont {A.~M.}\ \bibnamefont {Petersen}}, \bibinfo {author}
  {\bibfnamefont {F.}~\bibnamefont {Radicchi}}, \bibinfo {author}
  {\bibfnamefont {R.}~\bibnamefont {Sinatra}}, \bibinfo {author} {\bibfnamefont
  {B.}~\bibnamefont {Uzzi}},  \emph {et~al.},\ }\href@noop {} {\bibfield
  {journal} {\bibinfo  {journal} {Science}\ }\textbf {\bibinfo {volume}
  {359}},\ \bibinfo {pages} {eaao0185} (\bibinfo {year} {2018})}\BibitemShut
  {NoStop}%
\bibitem [{\citenamefont {Clauset}\ \emph {et~al.}(2017)\citenamefont
  {Clauset}, \citenamefont {Larremore},\ and\ \citenamefont
  {Sinatra}}]{clauset2017data}%
  \BibitemOpen
  \bibfield  {author} {\bibinfo {author} {\bibfnamefont {A.}~\bibnamefont
  {Clauset}}, \bibinfo {author} {\bibfnamefont {D.~B.}\ \bibnamefont
  {Larremore}}, \ and\ \bibinfo {author} {\bibfnamefont {R.}~\bibnamefont
  {Sinatra}},\ }\href@noop {} {\bibfield  {journal} {\bibinfo  {journal}
  {Science}\ }\textbf {\bibinfo {volume} {355}},\ \bibinfo {pages} {477}
  (\bibinfo {year} {2017})}\BibitemShut {NoStop}%
\bibitem [{\citenamefont {Sinatra}\ \emph {et~al.}(2015)\citenamefont
  {Sinatra}, \citenamefont {Deville}, \citenamefont {Szell}, \citenamefont
  {Wang},\ and\ \citenamefont {Barab{\'a}si}}]{sinatra2015century}%
  \BibitemOpen
  \bibfield  {author} {\bibinfo {author} {\bibfnamefont {R.}~\bibnamefont
  {Sinatra}}, \bibinfo {author} {\bibfnamefont {P.}~\bibnamefont {Deville}},
  \bibinfo {author} {\bibfnamefont {M.}~\bibnamefont {Szell}}, \bibinfo
  {author} {\bibfnamefont {D.}~\bibnamefont {Wang}}, \ and\ \bibinfo {author}
  {\bibfnamefont {A.-L.}\ \bibnamefont {Barab{\'a}si}},\ }\href@noop {}
  {\bibfield  {journal} {\bibinfo  {journal} {Nature Physics}\ }\textbf
  {\bibinfo {volume} {11}},\ \bibinfo {pages} {791} (\bibinfo {year}
  {2015})}\BibitemShut {NoStop}%
\bibitem [{\citenamefont {Battiston}\ \emph {et~al.}(2019)\citenamefont
  {Battiston}, \citenamefont {Musciotto}, \citenamefont {Wang}, \citenamefont
  {Barab{\'a}si}, \citenamefont {Szell},\ and\ \citenamefont
  {Sinatra}}]{battiston2019taking}%
  \BibitemOpen
  \bibfield  {author} {\bibinfo {author} {\bibfnamefont {F.}~\bibnamefont
  {Battiston}}, \bibinfo {author} {\bibfnamefont {F.}~\bibnamefont
  {Musciotto}}, \bibinfo {author} {\bibfnamefont {D.}~\bibnamefont {Wang}},
  \bibinfo {author} {\bibfnamefont {A.-L.}\ \bibnamefont {Barab{\'a}si}},
  \bibinfo {author} {\bibfnamefont {M.}~\bibnamefont {Szell}}, \ and\ \bibinfo
  {author} {\bibfnamefont {R.}~\bibnamefont {Sinatra}},\ }\href@noop {}
  {\bibfield  {journal} {\bibinfo  {journal} {Nature Reviews Physics}\ }\textbf
  {\bibinfo {volume} {1}},\ \bibinfo {pages} {89} (\bibinfo {year}
  {2019})}\BibitemShut {NoStop}%
\bibitem [{\citenamefont {Wagner}\ \emph {et~al.}(2011)\citenamefont {Wagner},
  \citenamefont {Roessner}, \citenamefont {Bobb}, \citenamefont {Klein},
  \citenamefont {Boyack}, \citenamefont {Keyton}, \citenamefont {Rafols},\ and\
  \citenamefont {B{\"o}rner}}]{53e9b58db7602d97040cf45d}%
  \BibitemOpen
  \bibfield  {author} {\bibinfo {author} {\bibfnamefont {C.~S.}\ \bibnamefont
  {Wagner}}, \bibinfo {author} {\bibfnamefont {J.~D.}\ \bibnamefont
  {Roessner}}, \bibinfo {author} {\bibfnamefont {K.}~\bibnamefont {Bobb}},
  \bibinfo {author} {\bibfnamefont {J.~T.}\ \bibnamefont {Klein}}, \bibinfo
  {author} {\bibfnamefont {K.~W.}\ \bibnamefont {Boyack}}, \bibinfo {author}
  {\bibfnamefont {J.}~\bibnamefont {Keyton}}, \bibinfo {author} {\bibfnamefont
  {I.}~\bibnamefont {Rafols}}, \ and\ \bibinfo {author} {\bibfnamefont
  {K.}~\bibnamefont {B{\"o}rner}},\ }\href@noop {} {\ \textbf {\bibinfo
  {volume} {5}},\ \bibinfo {pages} {14} (\bibinfo {year} {2011})}\BibitemShut
  {NoStop}%
\bibitem [{\citenamefont {Leydesdorff}\ and\ \citenamefont
  {Rafols}(2011)}]{RePEc:eee:infome:v:5:y:2011:i:1:p:87-100}%
  \BibitemOpen
  \bibfield  {author} {\bibinfo {author} {\bibfnamefont {L.}~\bibnamefont
  {Leydesdorff}}\ and\ \bibinfo {author} {\bibfnamefont {I.}~\bibnamefont
  {Rafols}},\ }\href {\doibase 10.1016/j.joi.2010.09.002} {\bibfield  {journal}
  {\bibinfo  {journal} {Journal of Informetrics}\ }\textbf {\bibinfo {volume}
  {5}},\ \bibinfo {pages} {87} (\bibinfo {year} {2011})}\BibitemShut {NoStop}%
\bibitem [{\citenamefont {Foster}\ \emph {et~al.}(2015)\citenamefont {Foster},
  \citenamefont {Rzhetsky},\ and\ \citenamefont
  {Evans}}]{doi:10.1177/0003122415601618}%
  \BibitemOpen
  \bibfield  {author} {\bibinfo {author} {\bibfnamefont {J.~G.}\ \bibnamefont
  {Foster}}, \bibinfo {author} {\bibfnamefont {A.}~\bibnamefont {Rzhetsky}}, \
  and\ \bibinfo {author} {\bibfnamefont {J.~A.}\ \bibnamefont {Evans}},\ }\href
  {\doibase 10.1177/0003122415601618} {\bibfield  {journal} {\bibinfo
  {journal} {American Sociological Review}\ }\textbf {\bibinfo {volume} {80}},\
  \bibinfo {pages} {875} (\bibinfo {year} {2015})},\ \Eprint
  {http://arxiv.org/abs/https://doi.org/10.1177/0003122415601618}
  {https://doi.org/10.1177/0003122415601618} \BibitemShut {NoStop}%
\bibitem [{\citenamefont {Aleta}\ \emph {et~al.}(2019)\citenamefont {Aleta},
  \citenamefont {Meloni}, \citenamefont {Perra},\ and\ \citenamefont
  {Moreno}}]{Aleta:2019aa}%
  \BibitemOpen
  \bibfield  {author} {\bibinfo {author} {\bibfnamefont {A.}~\bibnamefont
  {Aleta}}, \bibinfo {author} {\bibfnamefont {S.}~\bibnamefont {Meloni}},
  \bibinfo {author} {\bibfnamefont {N.}~\bibnamefont {Perra}}, \ and\ \bibinfo
  {author} {\bibfnamefont {Y.}~\bibnamefont {Moreno}},\ }\href {\doibase
  10.1140/epjds/s13688-019-0205-9} {\bibfield  {journal} {\bibinfo  {journal}
  {EPJ Data Science}\ }\textbf {\bibinfo {volume} {8}},\ \bibinfo {pages} {27}
  (\bibinfo {year} {2019})}\BibitemShut {NoStop}%
\bibitem [{\citenamefont {Bornmann}\ \emph {et~al.}(2011)\citenamefont
  {Bornmann}, \citenamefont {Leydesdorff}, \citenamefont {Walch-Solimena},\
  and\ \citenamefont {Ettl}}]{Bornmann2011}%
  \BibitemOpen
  \bibfield  {author} {\bibinfo {author} {\bibfnamefont {L.}~\bibnamefont
  {Bornmann}}, \bibinfo {author} {\bibfnamefont {L.}~\bibnamefont
  {Leydesdorff}}, \bibinfo {author} {\bibfnamefont {C.}~\bibnamefont
  {Walch-Solimena}}, \ and\ \bibinfo {author} {\bibfnamefont {C.}~\bibnamefont
  {Ettl}},\ }\href@noop {} {\bibfield  {journal} {\bibinfo  {journal} {Journal
  of Informetrics}\ }\textbf {\bibinfo {volume} {5}},\ \bibinfo {pages} {537}
  (\bibinfo {year} {2011})}\BibitemShut {NoStop}%
\bibitem [{\citenamefont {Zhang}\ \emph {et~al.}(2013)\citenamefont {Zhang},
  \citenamefont {Perra}, \citenamefont {Gon{\c{c}}alves}, \citenamefont
  {Ciulla},\ and\ \citenamefont {Vespignani}}]{zhang2013characterizing}%
  \BibitemOpen
  \bibfield  {author} {\bibinfo {author} {\bibfnamefont {Q.}~\bibnamefont
  {Zhang}}, \bibinfo {author} {\bibfnamefont {N.}~\bibnamefont {Perra}},
  \bibinfo {author} {\bibfnamefont {B.}~\bibnamefont {Gon{\c{c}}alves}},
  \bibinfo {author} {\bibfnamefont {F.}~\bibnamefont {Ciulla}}, \ and\ \bibinfo
  {author} {\bibfnamefont {A.}~\bibnamefont {Vespignani}},\ }\href@noop {}
  {\bibfield  {journal} {\bibinfo  {journal} {Scientific reports}\ }\textbf
  {\bibinfo {volume} {3}} (\bibinfo {year} {2013})}\BibitemShut {NoStop}%
\bibitem [{\citenamefont {Deville}\ \emph {et~al.}(2014)\citenamefont
  {Deville}, \citenamefont {Wang}, \citenamefont {Sinatra}, \citenamefont
  {Song}, \citenamefont {Blondel},\ and\ \citenamefont
  {Barab{\'a}si}}]{deville2014career}%
  \BibitemOpen
  \bibfield  {author} {\bibinfo {author} {\bibfnamefont {P.}~\bibnamefont
  {Deville}}, \bibinfo {author} {\bibfnamefont {D.}~\bibnamefont {Wang}},
  \bibinfo {author} {\bibfnamefont {R.}~\bibnamefont {Sinatra}}, \bibinfo
  {author} {\bibfnamefont {C.}~\bibnamefont {Song}}, \bibinfo {author}
  {\bibfnamefont {V.~D.}\ \bibnamefont {Blondel}}, \ and\ \bibinfo {author}
  {\bibfnamefont {A.-L.}\ \bibnamefont {Barab{\'a}si}},\ }\href@noop {}
  {\bibfield  {journal} {\bibinfo  {journal} {Scientific reports}\ }\textbf
  {\bibinfo {volume} {4}} (\bibinfo {year} {2014})}\BibitemShut {NoStop}%
\bibitem [{\citenamefont {Sinatra}\ \emph {et~al.}(2016)\citenamefont
  {Sinatra}, \citenamefont {Wang}, \citenamefont {Deville}, \citenamefont
  {Song},\ and\ \citenamefont {Barab{\'a}si}}]{sinatra2016quantifying}%
  \BibitemOpen
  \bibfield  {author} {\bibinfo {author} {\bibfnamefont {R.}~\bibnamefont
  {Sinatra}}, \bibinfo {author} {\bibfnamefont {D.}~\bibnamefont {Wang}},
  \bibinfo {author} {\bibfnamefont {P.}~\bibnamefont {Deville}}, \bibinfo
  {author} {\bibfnamefont {C.}~\bibnamefont {Song}}, \ and\ \bibinfo {author}
  {\bibfnamefont {A.-L.}\ \bibnamefont {Barab{\'a}si}},\ }\href@noop {}
  {\bibfield  {journal} {\bibinfo  {journal} {Science}\ }\textbf {\bibinfo
  {volume} {354}},\ \bibinfo {pages} {aaf5239} (\bibinfo {year}
  {2016})}\BibitemShut {NoStop}%
\bibitem [{\citenamefont {Newman}(2010)}]{newman10-1}%
  \BibitemOpen
  \bibfield  {author} {\bibinfo {author} {\bibfnamefont {M.}~\bibnamefont
  {Newman}},\ }\href@noop {} {\emph {\bibinfo {title} {Networks. An
  Introduction}}}\ (\bibinfo  {publisher} {Oxford University Press},\ \bibinfo
  {year} {2010})\BibitemShut {NoStop}%
\bibitem [{\citenamefont {Newman}(2001)}]{Newman2001Jan}%
  \BibitemOpen
  \bibfield  {author} {\bibinfo {author} {\bibfnamefont {M.~E.~J.}\
  \bibnamefont {Newman}},\ }\href {\doibase 10.1073/pnas.98.2.404} {\bibfield
  {journal} {\bibinfo  {journal} {Proc. Natl. Acad. Sci. U.S.A.}\ }\textbf
  {\bibinfo {volume} {98}},\ \bibinfo {pages} {404} (\bibinfo {year}
  {2001})}\BibitemShut {NoStop}%
\bibitem [{\citenamefont {Liben-Nowell}\ and\ \citenamefont
  {Kleinberg}(2003)}]{Nowell2003}%
  \BibitemOpen
  \bibfield  {author} {\bibinfo {author} {\bibfnamefont {D.}~\bibnamefont
  {Liben-Nowell}}\ and\ \bibinfo {author} {\bibfnamefont {J.}~\bibnamefont
  {Kleinberg}},\ }in\ \href {\doibase 10.1145/956863.956972} {\emph {\bibinfo
  {booktitle} {Proceedings of the Twelfth International Conference on
  Information and Knowledge Management}}},\ \bibinfo {series and number} {CIKM
  '03}\ (\bibinfo  {publisher} {Association for Computing Machinery},\ \bibinfo
  {address} {New York, NY, USA},\ \bibinfo {year} {2003})\ pp.\ \bibinfo
  {pages} {556--559}\BibitemShut {NoStop}%
\bibitem [{\citenamefont {Anand}\ \emph {et~al.}(2018)\citenamefont {Anand},
  \citenamefont {van Lelyveld}, \citenamefont {Banai}, \citenamefont
  {Friedrich}, \citenamefont {Garratt}, \citenamefont {Ha{\l}aj}, \citenamefont
  {Fique}, \citenamefont {Hansen}, \citenamefont {Jaramillo}, \citenamefont
  {Lee}, \citenamefont {Molina-Borboa}, \citenamefont {Nobili}, \citenamefont
  {Rajan}, \citenamefont {Salakhova}, \citenamefont {Silva}, \citenamefont
  {Silvestri},\ and\ \citenamefont {de~Souza}}]{Anand2018Apr}%
  \BibitemOpen
  \bibfield  {author} {\bibinfo {author} {\bibfnamefont {K.}~\bibnamefont
  {Anand}}, \bibinfo {author} {\bibfnamefont {I.}~\bibnamefont {van Lelyveld}},
  \bibinfo {author} {\bibfnamefont
  {{\ifmmode\acute{A}\else\'{A}\fi}.}~\bibnamefont {Banai}}, \bibinfo {author}
  {\bibfnamefont {S.}~\bibnamefont {Friedrich}}, \bibinfo {author}
  {\bibfnamefont {R.}~\bibnamefont {Garratt}}, \bibinfo {author} {\bibfnamefont
  {G.}~\bibnamefont {Ha{\l}aj}}, \bibinfo {author} {\bibfnamefont
  {J.}~\bibnamefont {Fique}}, \bibinfo {author} {\bibfnamefont
  {I.}~\bibnamefont {Hansen}}, \bibinfo {author} {\bibfnamefont {S.~M.}\
  \bibnamefont {Jaramillo}}, \bibinfo {author} {\bibfnamefont {H.}~\bibnamefont
  {Lee}}, \bibinfo {author} {\bibfnamefont {J.~L.}\ \bibnamefont
  {Molina-Borboa}}, \bibinfo {author} {\bibfnamefont {S.}~\bibnamefont
  {Nobili}}, \bibinfo {author} {\bibfnamefont {S.}~\bibnamefont {Rajan}},
  \bibinfo {author} {\bibfnamefont {D.}~\bibnamefont {Salakhova}}, \bibinfo
  {author} {\bibfnamefont {T.~C.}\ \bibnamefont {Silva}}, \bibinfo {author}
  {\bibfnamefont {L.}~\bibnamefont {Silvestri}}, \ and\ \bibinfo {author}
  {\bibfnamefont {S.~R.~S.}\ \bibnamefont {de~Souza}},\ }\href {\doibase
  10.1016/j.jfs.2017.05.012} {\bibfield  {journal} {\bibinfo  {journal}
  {Journal of Financial Stability}\ }\textbf {\bibinfo {volume} {35}},\
  \bibinfo {pages} {107} (\bibinfo {year} {2018})}\BibitemShut {NoStop}%
\bibitem [{\citenamefont {Wang}\ \emph {et~al.}(2015)\citenamefont {Wang},
  \citenamefont {Xu}, \citenamefont {Wu},\ and\ \citenamefont
  {Zhou}}]{Wang2015Jan}%
  \BibitemOpen
  \bibfield  {author} {\bibinfo {author} {\bibfnamefont {P.}~\bibnamefont
  {Wang}}, \bibinfo {author} {\bibfnamefont {B.}~\bibnamefont {Xu}}, \bibinfo
  {author} {\bibfnamefont {Y.}~\bibnamefont {Wu}}, \ and\ \bibinfo {author}
  {\bibfnamefont {X.}~\bibnamefont {Zhou}},\ }\href {\doibase
  10.1007/s11432-014-5237-y} {\bibfield  {journal} {\bibinfo  {journal} {Sci.
  China Inf. Sci.}\ }\textbf {\bibinfo {volume} {58}},\ \bibinfo {pages} {1}
  (\bibinfo {year} {2015})}\BibitemShut {NoStop}%
\bibitem [{\citenamefont {Marjan}\ \emph {et~al.}(2018)\citenamefont {Marjan},
  \citenamefont {Zaki},\ and\ \citenamefont {Mohamed}}]{Marjan2018Oct}%
  \BibitemOpen
  \bibfield  {author} {\bibinfo {author} {\bibfnamefont {M.}~\bibnamefont
  {Marjan}}, \bibinfo {author} {\bibfnamefont {N.}~\bibnamefont {Zaki}}, \ and\
  \bibinfo {author} {\bibfnamefont {E.~A.}\ \bibnamefont {Mohamed}},\ }\href
  {\doibase 10.1109/CIST.2018.8596511} {\bibfield  {journal} {\bibinfo
  {journal} {2018 IEEE 5th International Congress on Information Science and
  Technology (CiSt)}\ ,\ \bibinfo {pages} {200}} (\bibinfo {year}
  {2018})}\BibitemShut {NoStop}%
\bibitem [{\citenamefont
  {Mart{\ifmmode\acute{\imath}\else\'{\i}\fi}nezV{\ifmmode\acute{\imath}\else\'{\i}\fi}ctor}\
  \emph {et~al.}(2016)\citenamefont
  {Mart{\ifmmode\acute{\imath}\else\'{\i}\fi}nezV{\ifmmode\acute{\imath}\else\'{\i}\fi}ctor},
  \citenamefont {BerzalFernando},\ and\ \citenamefont
  {CuberoJuan-Carlos}}]{MartnnezVnctor2016Dec}%
  \BibitemOpen
  \bibfield  {author} {\bibinfo {author} {\bibnamefont
  {Mart{\ifmmode\acute{\imath}\else\'{\i}\fi}nezV{\ifmmode\acute{\imath}\else\'{\i}\fi}ctor}},
  \bibinfo {author} {\bibnamefont {BerzalFernando}}, \ and\ \bibinfo {author}
  {\bibnamefont {CuberoJuan-Carlos}},\ }\href
  {https://dl.acm.org/doi/10.1145/3012704} {\bibfield  {journal} {\bibinfo
  {journal} {ACM Comput. Surv.}\ } (\bibinfo {year} {2016})}\BibitemShut
  {NoStop}%
\bibitem [{\citenamefont {Bliss}\ \emph {et~al.}(2014)\citenamefont {Bliss},
  \citenamefont {Frank}, \citenamefont {Danforth},\ and\ \citenamefont
  {Dodds}}]{Bliss2014Sep}%
  \BibitemOpen
  \bibfield  {author} {\bibinfo {author} {\bibfnamefont {C.~A.}\ \bibnamefont
  {Bliss}}, \bibinfo {author} {\bibfnamefont {M.~R.}\ \bibnamefont {Frank}},
  \bibinfo {author} {\bibfnamefont {C.~M.}\ \bibnamefont {Danforth}}, \ and\
  \bibinfo {author} {\bibfnamefont {P.~S.}\ \bibnamefont {Dodds}},\ }\href
  {\doibase 10.1016/j.jocs.2014.01.003} {\bibfield  {journal} {\bibinfo
  {journal} {Journal of Computational Science}\ }\textbf {\bibinfo {volume}
  {5}},\ \bibinfo {pages} {750} (\bibinfo {year} {2014})}\BibitemShut {NoStop}%
\bibitem [{\citenamefont {Adamic}\ and\ \citenamefont
  {Adar}(2003)}]{Adamic2003Jul}%
  \BibitemOpen
  \bibfield  {author} {\bibinfo {author} {\bibfnamefont {L.~A.}\ \bibnamefont
  {Adamic}}\ and\ \bibinfo {author} {\bibfnamefont {E.}~\bibnamefont {Adar}},\
  }\href {\doibase 10.1016/S0378-8733(03)00009-1} {\bibfield  {journal}
  {\bibinfo  {journal} {Social Networks}\ }\textbf {\bibinfo {volume} {25}},\
  \bibinfo {pages} {211} (\bibinfo {year} {2003})}\BibitemShut {NoStop}%
\bibitem [{\citenamefont {Clauset}\ \emph {et~al.}(2008)\citenamefont
  {Clauset}, \citenamefont {Moore},\ and\ \citenamefont
  {Newman}}]{Clauset2008May}%
  \BibitemOpen
  \bibfield  {author} {\bibinfo {author} {\bibfnamefont {A.}~\bibnamefont
  {Clauset}}, \bibinfo {author} {\bibfnamefont {C.}~\bibnamefont {Moore}}, \
  and\ \bibinfo {author} {\bibfnamefont {M.~E.~J.}\ \bibnamefont {Newman}},\
  }\href {\doibase 10.1038/nature06830} {\bibfield  {journal} {\bibinfo
  {journal} {Nature}\ }\textbf {\bibinfo {volume} {453}},\ \bibinfo {pages}
  {98} (\bibinfo {year} {2008})}\BibitemShut {NoStop}%
\bibitem [{\citenamefont {Cho}\ and\ \citenamefont {Yu}(2018)}]{Cho2018Dec}%
  \BibitemOpen
  \bibfield  {author} {\bibinfo {author} {\bibfnamefont {H.}~\bibnamefont
  {Cho}}\ and\ \bibinfo {author} {\bibfnamefont {Y.}~\bibnamefont {Yu}},\
  }\href {\doibase 10.1007/s13278-018-0501-6} {\bibfield  {journal} {\bibinfo
  {journal} {Soc. Netw. Anal. Min.}\ }\textbf {\bibinfo {volume} {8}},\
  \bibinfo {pages} {1} (\bibinfo {year} {2018})}\BibitemShut {NoStop}%
\bibitem [{\citenamefont {Al~Hasan}\ \emph {et~al.}(2006)\citenamefont
  {Al~Hasan}, \citenamefont {Chaoji}, \citenamefont {Salem},\ and\
  \citenamefont {Zaki}}]{Hassan2006}%
  \BibitemOpen
  \bibfield  {author} {\bibinfo {author} {\bibfnamefont {M.}~\bibnamefont
  {Al~Hasan}}, \bibinfo {author} {\bibfnamefont {V.}~\bibnamefont {Chaoji}},
  \bibinfo {author} {\bibfnamefont {S.}~\bibnamefont {Salem}}, \ and\ \bibinfo
  {author} {\bibfnamefont {M.}~\bibnamefont {Zaki}},\ }in\ \href@noop {} {\emph
  {\bibinfo {booktitle} {SDM06: workshop on link analysis, counter-terrorism
  and security}}},\ Vol.~\bibinfo {volume} {30}\ (\bibinfo {year} {2006})\ pp.\
  \bibinfo {pages} {798--805}\BibitemShut {NoStop}%
\bibitem [{\citenamefont {Moradabadi}\ and\ \citenamefont
  {Meybodi}(2017)}]{Moradabadi2017Sep}%
  \BibitemOpen
  \bibfield  {author} {\bibinfo {author} {\bibfnamefont {B.}~\bibnamefont
  {Moradabadi}}\ and\ \bibinfo {author} {\bibfnamefont {M.~R.}\ \bibnamefont
  {Meybodi}},\ }\href {\doibase 10.1016/j.physa.2017.04.019} {\bibfield
  {journal} {\bibinfo  {journal} {Physica A}\ }\textbf {\bibinfo {volume}
  {482}},\ \bibinfo {pages} {422} (\bibinfo {year} {2017})}\BibitemShut
  {NoStop}%
\bibitem [{\citenamefont {Najari}\ \emph {et~al.}(2019)\citenamefont {Najari},
  \citenamefont {Salehi}, \citenamefont {Ranjbar},\ and\ \citenamefont
  {Jalili}}]{Najari2019Dec}%
  \BibitemOpen
  \bibfield  {author} {\bibinfo {author} {\bibfnamefont {S.}~\bibnamefont
  {Najari}}, \bibinfo {author} {\bibfnamefont {M.}~\bibnamefont {Salehi}},
  \bibinfo {author} {\bibfnamefont {V.}~\bibnamefont {Ranjbar}}, \ and\
  \bibinfo {author} {\bibfnamefont {M.}~\bibnamefont {Jalili}},\ }\href
  {\doibase 10.1016/j.physa.2019.04.214} {\bibfield  {journal} {\bibinfo
  {journal} {Physica A}\ }\textbf {\bibinfo {volume} {536}},\ \bibinfo {pages}
  {120978} (\bibinfo {year} {2019})}\BibitemShut {NoStop}%
\bibitem [{\citenamefont {Sett}\ \emph {et~al.}(2016)\citenamefont {Sett},
  \citenamefont {Ranbir~Singh},\ and\ \citenamefont {Nandi}}]{Sett2016Jan}%
  \BibitemOpen
  \bibfield  {author} {\bibinfo {author} {\bibfnamefont {N.}~\bibnamefont
  {Sett}}, \bibinfo {author} {\bibfnamefont {S.}~\bibnamefont {Ranbir~Singh}},
  \ and\ \bibinfo {author} {\bibfnamefont {S.}~\bibnamefont {Nandi}},\ }\href
  {\doibase 10.1016/j.neucom.2014.11.089} {\bibfield  {journal} {\bibinfo
  {journal} {Neurocomputing}\ }\textbf {\bibinfo {volume} {172}},\ \bibinfo
  {pages} {71} (\bibinfo {year} {2016})}\BibitemShut {NoStop}%
\bibitem [{\citenamefont {ZhangJinzhu}(2017)}]{ZhangJinzhu2017Jan}%
  \BibitemOpen
  \bibfield  {author} {\bibinfo {author} {\bibnamefont {ZhangJinzhu}},\ }\href
  {https://dl.acm.org/doi/10.1016/j.ipm.2016.06.005} {\bibfield  {journal}
  {\bibinfo  {journal} {Information Processing and Management: an International
  Journal}\ } (\bibinfo {year} {2017})}\BibitemShut {NoStop}%
\bibitem [{\citenamefont {Jia}\ \emph {et~al.}(2020)\citenamefont {Jia},
  \citenamefont {Ran},\ and\ \citenamefont {Xu}}]{Jia2020}%
  \BibitemOpen
  \bibfield  {author} {\bibinfo {author} {\bibfnamefont {T.}~\bibnamefont
  {Jia}}, \bibinfo {author} {\bibfnamefont {Y.}~\bibnamefont {Ran}}, \ and\
  \bibinfo {author} {\bibfnamefont {X.}~\bibnamefont {Xu}},\ }in\ \href@noop {}
  {\emph {\bibinfo {booktitle} {NetSci-X 2020}}}\ (\bibinfo {year}
  {2020})\BibitemShut {NoStop}%
\bibitem [{\citenamefont {Kivel{\ifmmode\ddot{a}\else\"{a}\fi}}\ \emph
  {et~al.}(2014)\citenamefont {Kivel{\ifmmode\ddot{a}\else\"{a}\fi}},
  \citenamefont {Arenas}, \citenamefont {Barthelemy}, \citenamefont {Gleeson},
  \citenamefont {Moreno},\ and\ \citenamefont {Porter}}]{Kivela2014Sep}%
  \BibitemOpen
  \bibfield  {author} {\bibinfo {author} {\bibfnamefont {M.}~\bibnamefont
  {Kivel{\ifmmode\ddot{a}\else\"{a}\fi}}}, \bibinfo {author} {\bibfnamefont
  {A.}~\bibnamefont {Arenas}}, \bibinfo {author} {\bibfnamefont
  {M.}~\bibnamefont {Barthelemy}}, \bibinfo {author} {\bibfnamefont {J.~P.}\
  \bibnamefont {Gleeson}}, \bibinfo {author} {\bibfnamefont {Y.}~\bibnamefont
  {Moreno}}, \ and\ \bibinfo {author} {\bibfnamefont {M.~A.}\ \bibnamefont
  {Porter}},\ }\href {\doibase 10.1093/comnet/cnu016} {\bibfield  {journal}
  {\bibinfo  {journal} {J. Complex Networks}\ }\textbf {\bibinfo {volume}
  {2}},\ \bibinfo {pages} {203} (\bibinfo {year} {2014})}\BibitemShut {NoStop}%
\bibitem [{\citenamefont {Aleta}\ and\ \citenamefont
  {Moreno}(2019)}]{Aleta2019Mar}%
  \BibitemOpen
  \bibfield  {author} {\bibinfo {author} {\bibfnamefont {A.}~\bibnamefont
  {Aleta}}\ and\ \bibinfo {author} {\bibfnamefont {Y.}~\bibnamefont {Moreno}},\
  }\href {\doibase 10.1146/annurev-conmatphys-031218-013259} {\bibfield
  {journal} {\bibinfo  {journal} {Annu. Rev. Condens. Matter Phys.}\ }\textbf
  {\bibinfo {volume} {10}},\ \bibinfo {pages} {45} (\bibinfo {year}
  {2019})}\BibitemShut {NoStop}%
\bibitem [{\citenamefont {Radicchi}\ \emph {et~al.}(2009)\citenamefont
  {Radicchi}, \citenamefont {Fortunato}, \citenamefont {Markines},\ and\
  \citenamefont {Vespignani}}]{Radicchi2009Nov}%
  \BibitemOpen
  \bibfield  {author} {\bibinfo {author} {\bibfnamefont {F.}~\bibnamefont
  {Radicchi}}, \bibinfo {author} {\bibfnamefont {S.}~\bibnamefont {Fortunato}},
  \bibinfo {author} {\bibfnamefont {B.}~\bibnamefont {Markines}}, \ and\
  \bibinfo {author} {\bibfnamefont {A.}~\bibnamefont {Vespignani}},\ }\href
  {\doibase 10.1103/PhysRevE.80.056103} {\bibfield  {journal} {\bibinfo
  {journal} {Phys. Rev. E}\ }\textbf {\bibinfo {volume} {80}},\ \bibinfo
  {pages} {056103} (\bibinfo {year} {2009})}\BibitemShut {NoStop}%
\bibitem [{\citenamefont {Tang}\ \emph {et~al.}(2008)\citenamefont {Tang},
  \citenamefont {Zhang}, \citenamefont {Yao}, \citenamefont {Li}, \citenamefont
  {Zhang},\ and\ \citenamefont {Su}}]{OAG1}%
  \BibitemOpen
  \bibfield  {author} {\bibinfo {author} {\bibfnamefont {J.}~\bibnamefont
  {Tang}}, \bibinfo {author} {\bibfnamefont {J.}~\bibnamefont {Zhang}},
  \bibinfo {author} {\bibfnamefont {L.}~\bibnamefont {Yao}}, \bibinfo {author}
  {\bibfnamefont {J.}~\bibnamefont {Li}}, \bibinfo {author} {\bibfnamefont
  {L.}~\bibnamefont {Zhang}}, \ and\ \bibinfo {author} {\bibfnamefont
  {Z.}~\bibnamefont {Su}},\ }in\ \href@noop {} {\emph {\bibinfo {booktitle}
  {Proceedings of the Fourteenth ACM SIGKDD International Conference on
  Knowledge Discovery and Data Mining (SIGKDD'2008)}}}\ (\bibinfo {year}
  {2008})\ pp.\ \bibinfo {pages} {990--998}\BibitemShut {NoStop}%
\bibitem [{\citenamefont {Menichetti}\ \emph {et~al.}(2014)\citenamefont
  {Menichetti}, \citenamefont {Remondini}, \citenamefont {Panzarasa},
  \citenamefont {Mondrag{\'o}n},\ and\ \citenamefont
  {Bianconi}}]{10.1371/journal.pone.0097857}%
  \BibitemOpen
  \bibfield  {author} {\bibinfo {author} {\bibfnamefont {G.}~\bibnamefont
  {Menichetti}}, \bibinfo {author} {\bibfnamefont {D.}~\bibnamefont
  {Remondini}}, \bibinfo {author} {\bibfnamefont {P.}~\bibnamefont
  {Panzarasa}}, \bibinfo {author} {\bibfnamefont {R.~J.}\ \bibnamefont
  {Mondrag{\'o}n}}, \ and\ \bibinfo {author} {\bibfnamefont {G.}~\bibnamefont
  {Bianconi}},\ }\href {\doibase 10.1371/journal.pone.0097857} {\bibfield
  {journal} {\bibinfo  {journal} {PLOS ONE}\ }\textbf {\bibinfo {volume} {9}},\
  \bibinfo {pages} {1} (\bibinfo {year} {2014})}\BibitemShut {NoStop}%
\bibitem [{\citenamefont {Battiston}\ \emph {et~al.}(2014)\citenamefont
  {Battiston}, \citenamefont {Nicosia},\ and\ \citenamefont
  {Latora}}]{PhysRevE.89.032804}%
  \BibitemOpen
  \bibfield  {author} {\bibinfo {author} {\bibfnamefont {F.}~\bibnamefont
  {Battiston}}, \bibinfo {author} {\bibfnamefont {V.}~\bibnamefont {Nicosia}},
  \ and\ \bibinfo {author} {\bibfnamefont {V.}~\bibnamefont {Latora}},\ }\href
  {\doibase 10.1103/PhysRevE.89.032804} {\bibfield  {journal} {\bibinfo
  {journal} {Phys. Rev. E}\ }\textbf {\bibinfo {volume} {89}},\ \bibinfo
  {pages} {032804} (\bibinfo {year} {2014})}\BibitemShut {NoStop}%
\bibitem [{\citenamefont {Cozzo}\ \emph {et~al.}(2015)\citenamefont {Cozzo},
  \citenamefont {Kivel{\ifmmode\ddot{a}\else\"{a}\fi}}, \citenamefont
  {De~Domenico}, \citenamefont {Sol{\ifmmode\acute{e}\else\'{e}\fi}-Ribalta},
  \citenamefont {Arenas}, \citenamefont {G{\ifmmode\acute{o}\else\'{o}\fi}mez},
  \citenamefont {Porter},\ and\ \citenamefont {Moreno}}]{Cozzo2015Jul}%
  \BibitemOpen
  \bibfield  {author} {\bibinfo {author} {\bibfnamefont {E.}~\bibnamefont
  {Cozzo}}, \bibinfo {author} {\bibfnamefont {M.}~\bibnamefont
  {Kivel{\ifmmode\ddot{a}\else\"{a}\fi}}}, \bibinfo {author} {\bibfnamefont
  {M.}~\bibnamefont {De~Domenico}}, \bibinfo {author} {\bibfnamefont
  {A.}~\bibnamefont {Sol{\ifmmode\acute{e}\else\'{e}\fi}-Ribalta}}, \bibinfo
  {author} {\bibfnamefont {A.}~\bibnamefont {Arenas}}, \bibinfo {author}
  {\bibfnamefont {S.}~\bibnamefont {G{\ifmmode\acute{o}\else\'{o}\fi}mez}},
  \bibinfo {author} {\bibfnamefont {M.~A.}\ \bibnamefont {Porter}}, \ and\
  \bibinfo {author} {\bibfnamefont {Y.}~\bibnamefont {Moreno}},\ }\href
  {\doibase 10.1088/1367-2630/17/7/073029} {\bibfield  {journal} {\bibinfo
  {journal} {New J. Phys.}\ }\textbf {\bibinfo {volume} {17}},\ \bibinfo
  {pages} {073029} (\bibinfo {year} {2015})}\BibitemShut {NoStop}%
\bibitem [{\citenamefont {Davis}\ \emph {et~al.}(2013)\citenamefont {Davis},
  \citenamefont {Lichtenwalter},\ and\ \citenamefont {Chawla}}]{Davis2013Jun}%
  \BibitemOpen
  \bibfield  {author} {\bibinfo {author} {\bibfnamefont {D.}~\bibnamefont
  {Davis}}, \bibinfo {author} {\bibfnamefont {R.}~\bibnamefont
  {Lichtenwalter}}, \ and\ \bibinfo {author} {\bibfnamefont {N.~V.}\
  \bibnamefont {Chawla}},\ }\href {\doibase 10.1007/s13278-012-0068-6}
  {\bibfield  {journal} {\bibinfo  {journal} {Soc. Netw. Anal. Min.}\ }\textbf
  {\bibinfo {volume} {3}},\ \bibinfo {pages} {127} (\bibinfo {year}
  {2013})}\BibitemShut {NoStop}%
\bibitem [{\citenamefont {Kleineberg}\ \emph {et~al.}(2016)\citenamefont
  {Kleineberg}, \citenamefont
  {Bogu{\ifmmode\tilde{n}\else\~{n}\fi}{\ifmmode\acute{a}\else\'{a}\fi}},
  \citenamefont {{\ifmmode\acute{A}\else\'{A}\fi}ngeles Serrano},\ and\
  \citenamefont {Papadopoulos}}]{Kleineberg2016Jul}%
  \BibitemOpen
  \bibfield  {author} {\bibinfo {author} {\bibfnamefont {K.-K.}\ \bibnamefont
  {Kleineberg}}, \bibinfo {author} {\bibfnamefont {M.}~\bibnamefont
  {Bogu{\ifmmode\tilde{n}\else\~{n}\fi}{\ifmmode\acute{a}\else\'{a}\fi}}},
  \bibinfo {author} {\bibfnamefont {M.}~\bibnamefont
  {{\ifmmode\acute{A}\else\'{A}\fi}ngeles Serrano}}, \ and\ \bibinfo {author}
  {\bibfnamefont {F.}~\bibnamefont {Papadopoulos}},\ }\href {\doibase
  10.1038/nphys3812} {\bibfield  {journal} {\bibinfo  {journal} {Nat. Phys.}\
  }\textbf {\bibinfo {volume} {12}},\ \bibinfo {pages} {1076} (\bibinfo {year}
  {2016})}\BibitemShut {NoStop}%
\bibitem [{\citenamefont {Lu}\ \emph {et~al.}(2010)\citenamefont {Lu},
  \citenamefont {Savas}, \citenamefont {Tang},\ and\ \citenamefont
  {Dhillon}}]{Lu2010Dec}%
  \BibitemOpen
  \bibfield  {author} {\bibinfo {author} {\bibfnamefont {Z.}~\bibnamefont
  {Lu}}, \bibinfo {author} {\bibfnamefont {B.}~\bibnamefont {Savas}}, \bibinfo
  {author} {\bibfnamefont {W.}~\bibnamefont {Tang}}, \ and\ \bibinfo {author}
  {\bibfnamefont {I.~S.}\ \bibnamefont {Dhillon}},\ }\href {\doibase
  10.1109/ICDM.2010.112} {\bibfield  {journal} {\bibinfo  {journal} {2010 IEEE
  International Conference on Data Mining}\ ,\ \bibinfo {pages} {923}}
  (\bibinfo {year} {2010})}\BibitemShut {NoStop}%
\bibitem [{\citenamefont {Matsuno}\ and\ \citenamefont
  {Murata}(2018)}]{Matsuno2018}%
  \BibitemOpen
  \bibfield  {author} {\bibinfo {author} {\bibfnamefont {R.}~\bibnamefont
  {Matsuno}}\ and\ \bibinfo {author} {\bibfnamefont {T.}~\bibnamefont
  {Murata}},\ }in\ \href {\doibase 10.1145/3184558.3191565} {\emph {\bibinfo
  {booktitle} {Companion Proceedings of the The Web Conference 2018}}},\
  \bibinfo {series and number} {WWW '18}\ (\bibinfo  {publisher} {International
  World Wide Web Conferences Steering Committee},\ \bibinfo {address} {Republic
  and Canton of Geneva, CHE},\ \bibinfo {year} {2018})\ pp.\ \bibinfo {pages}
  {1261--1268}\BibitemShut {NoStop}%
\bibitem [{\citenamefont {Pujari}\ and\ \citenamefont
  {Kanawati}(2015)}]{Pujari2015}%
  \BibitemOpen
  \bibfield  {author} {\bibinfo {author} {\bibfnamefont {M.}~\bibnamefont
  {Pujari}}\ and\ \bibinfo {author} {\bibfnamefont {R.}~\bibnamefont
  {Kanawati}},\ }\href {\doibase 10.3934/nhm.2015.10.17} {\bibfield  {journal}
  {\bibinfo  {journal} {Networks {\&} Heterogeneous Media}\ }\textbf {\bibinfo
  {volume} {10}},\ \bibinfo {pages} {17} (\bibinfo {year} {2015})}\BibitemShut
  {NoStop}%
\bibitem [{\citenamefont {Hajibagheri}\ \emph {et~al.}(2016)\citenamefont
  {Hajibagheri}, \citenamefont {Sukthankar},\ and\ \citenamefont
  {Lakkaraju}}]{Hajibagheri2016Aug}%
  \BibitemOpen
  \bibfield  {author} {\bibinfo {author} {\bibfnamefont {A.}~\bibnamefont
  {Hajibagheri}}, \bibinfo {author} {\bibfnamefont {G.}~\bibnamefont
  {Sukthankar}}, \ and\ \bibinfo {author} {\bibfnamefont {K.}~\bibnamefont
  {Lakkaraju}},\ }\href {\doibase 10.1109/ASONAM.2016.7752375} {\bibfield
  {journal} {\bibinfo  {journal} {2016 IEEE/ACM International Conference on
  Advances in Social Networks Analysis and Mining (ASONAM)}\ ,\ \bibinfo
  {pages} {1079}} (\bibinfo {year} {2016})}\BibitemShut {NoStop}%
\bibitem [{\citenamefont {Yao}\ \emph {et~al.}(2016)\citenamefont {Yao},
  \citenamefont {Wang}, \citenamefont {Pan},\ and\ \citenamefont
  {Yao}}]{Yao2016Jan}%
  \BibitemOpen
  \bibfield  {author} {\bibinfo {author} {\bibfnamefont {L.}~\bibnamefont
  {Yao}}, \bibinfo {author} {\bibfnamefont {L.}~\bibnamefont {Wang}}, \bibinfo
  {author} {\bibfnamefont {L.}~\bibnamefont {Pan}}, \ and\ \bibinfo {author}
  {\bibfnamefont {K.}~\bibnamefont {Yao}},\ }\href {\doibase
  10.1016/j.procs.2016.04.102} {\bibfield  {journal} {\bibinfo  {journal}
  {Procedia Comput. Sci.}\ }\textbf {\bibinfo {volume} {83}},\ \bibinfo {pages}
  {82} (\bibinfo {year} {2016})}\BibitemShut {NoStop}%
\bibitem [{\citenamefont {Jalili}\ \emph {et~al.}(2016)\citenamefont {Jalili},
  \citenamefont {Orouskhani}, \citenamefont {Asgari}, \citenamefont
  {Alipourfard},\ and\ \citenamefont {Perc}}]{Jalili2016Nov}%
  \BibitemOpen
  \bibfield  {author} {\bibinfo {author} {\bibfnamefont {M.}~\bibnamefont
  {Jalili}}, \bibinfo {author} {\bibfnamefont {Y.}~\bibnamefont {Orouskhani}},
  \bibinfo {author} {\bibfnamefont {M.}~\bibnamefont {Asgari}}, \bibinfo
  {author} {\bibfnamefont {N.}~\bibnamefont {Alipourfard}}, \ and\ \bibinfo
  {author} {\bibfnamefont {M.}~\bibnamefont {Perc}},\ }\href {\doibase
  10.1098/rsos.160863} {\bibfield  {journal} {\bibinfo  {journal} {R. Soc. Open
  Sci.}\ } (\bibinfo {year} {2016}),\ 10.1098/rsos.160863}\BibitemShut
  {NoStop}%
\bibitem [{\citenamefont {Hristova}\ \emph {et~al.}(2016)\citenamefont
  {Hristova}, \citenamefont {Noulas}, \citenamefont {Brown}, \citenamefont
  {Musolesi},\ and\ \citenamefont {Mascolo}}]{Hristova2016Dec}%
  \BibitemOpen
  \bibfield  {author} {\bibinfo {author} {\bibfnamefont {D.}~\bibnamefont
  {Hristova}}, \bibinfo {author} {\bibfnamefont {A.}~\bibnamefont {Noulas}},
  \bibinfo {author} {\bibfnamefont {C.}~\bibnamefont {Brown}}, \bibinfo
  {author} {\bibfnamefont {M.}~\bibnamefont {Musolesi}}, \ and\ \bibinfo
  {author} {\bibfnamefont {C.}~\bibnamefont {Mascolo}},\ }\href {\doibase
  10.1140/epjds/s13688-016-0087-z} {\bibfield  {journal} {\bibinfo  {journal}
  {EPJ Data Sci.}\ }\textbf {\bibinfo {volume} {5}},\ \bibinfo {pages} {1}
  (\bibinfo {year} {2016})}\BibitemShut {NoStop}%
\bibitem [{\citenamefont {Mandal}\ \emph {et~al.}(2018)\citenamefont {Mandal},
  \citenamefont {Mirchev}, \citenamefont {Gramatikov},\ and\ \citenamefont
  {Mishkovski}}]{Mandal2018Nov}%
  \BibitemOpen
  \bibfield  {author} {\bibinfo {author} {\bibfnamefont {H.}~\bibnamefont
  {Mandal}}, \bibinfo {author} {\bibfnamefont {M.}~\bibnamefont {Mirchev}},
  \bibinfo {author} {\bibfnamefont {S.}~\bibnamefont {Gramatikov}}, \ and\
  \bibinfo {author} {\bibfnamefont {I.}~\bibnamefont {Mishkovski}},\ }\href
  {\doibase 10.1109/TELFOR.2018.8612122} {\bibfield  {journal} {\bibinfo
  {journal} {2018 26th Telecommunications Forum (TELFOR)}\ ,\ \bibinfo {pages}
  {1}} (\bibinfo {year} {2018})}\BibitemShut {NoStop}%
\bibitem [{\citenamefont {Junuthula}\ \emph {et~al.}(2018)\citenamefont
  {Junuthula}, \citenamefont {Xu},\ and\ \citenamefont
  {Devabhaktuni}}]{Junuthula2018}%
  \BibitemOpen
  \bibfield  {author} {\bibinfo {author} {\bibfnamefont {R.}~\bibnamefont
  {Junuthula}}, \bibinfo {author} {\bibfnamefont {K.}~\bibnamefont {Xu}}, \
  and\ \bibinfo {author} {\bibfnamefont {V.}~\bibnamefont {Devabhaktuni}},\
  }\href@noop {} {\  (\bibinfo {year} {2018})}\BibitemShut {NoStop}%
\bibitem [{\citenamefont {Sharma}\ and\ \citenamefont
  {Singh}(2015)}]{Sharma2015Nov}%
  \BibitemOpen
  \bibfield  {author} {\bibinfo {author} {\bibfnamefont {S.}~\bibnamefont
  {Sharma}}\ and\ \bibinfo {author} {\bibfnamefont {A.}~\bibnamefont {Singh}},\
  }\href {\doibase 10.1109/SITIS.2015.93} {\emph {\bibinfo {title} {{An
  Efficient Method for Link Prediction in Complex Multiplex Networks}}}}\
  (\bibinfo  {publisher} {IEEE},\ \bibinfo {year} {2015})\BibitemShut {NoStop}%
\bibitem [{\citenamefont {Yao}\ \emph {et~al.}(2017)\citenamefont {Yao},
  \citenamefont {Zhang}, \citenamefont {Yang}, \citenamefont {Yuan},
  \citenamefont {Sun}, \citenamefont {Qiu},\ and\ \citenamefont
  {Hu}}]{Yao2017Jul}%
  \BibitemOpen
  \bibfield  {author} {\bibinfo {author} {\bibfnamefont {Y.}~\bibnamefont
  {Yao}}, \bibinfo {author} {\bibfnamefont {R.}~\bibnamefont {Zhang}}, \bibinfo
  {author} {\bibfnamefont {F.}~\bibnamefont {Yang}}, \bibinfo {author}
  {\bibfnamefont {Y.}~\bibnamefont {Yuan}}, \bibinfo {author} {\bibfnamefont
  {Q.}~\bibnamefont {Sun}}, \bibinfo {author} {\bibfnamefont {Y.}~\bibnamefont
  {Qiu}}, \ and\ \bibinfo {author} {\bibfnamefont {R.}~\bibnamefont {Hu}},\
  }\href {\doibase 10.1142/S0129183117501017} {\bibfield  {journal} {\bibinfo
  {journal} {Int. J. Mod. Phys. C}\ }\textbf {\bibinfo {volume} {28}},\
  \bibinfo {pages} {1750101} (\bibinfo {year} {2017})}\BibitemShut {NoStop}%
\bibitem [{\citenamefont {Samei}\ and\ \citenamefont
  {Jalili}(2019)}]{Samei2019Oct}%
  \BibitemOpen
  \bibfield  {author} {\bibinfo {author} {\bibfnamefont {Z.}~\bibnamefont
  {Samei}}\ and\ \bibinfo {author} {\bibfnamefont {M.}~\bibnamefont {Jalili}},\
  }\href {\doibase 10.1093/comnet/cnz007} {\bibfield  {journal} {\bibinfo
  {journal} {J. Complex Networks}\ }\textbf {\bibinfo {volume} {7}},\ \bibinfo
  {pages} {641} (\bibinfo {year} {2019})}\BibitemShut {NoStop}%
\bibitem [{\citenamefont {Starnini}\ \emph {et~al.}(2017)\citenamefont
  {Starnini}, \citenamefont {Baronchelli},\ and\ \citenamefont
  {Pastor-Satorras}}]{Starnini:2017aa}%
  \BibitemOpen
  \bibfield  {author} {\bibinfo {author} {\bibfnamefont {M.}~\bibnamefont
  {Starnini}}, \bibinfo {author} {\bibfnamefont {A.}~\bibnamefont
  {Baronchelli}}, \ and\ \bibinfo {author} {\bibfnamefont {R.}~\bibnamefont
  {Pastor-Satorras}},\ }\href {\doibase 10.1038/s41598-017-07591-0} {\bibfield
  {journal} {\bibinfo  {journal} {Scientific Reports}\ }\textbf {\bibinfo
  {volume} {7}},\ \bibinfo {pages} {8597} (\bibinfo {year} {2017})}\BibitemShut
  {NoStop}%
\bibitem [{\citenamefont {Muschelli}(2019)}]{Muschelli2019Dec}%
  \BibitemOpen
  \bibfield  {author} {\bibinfo {author} {\bibfnamefont {J.}~\bibnamefont
  {Muschelli}},\ }\href {\doibase 10.1007/s00357-019-09345-1} {\bibfield
  {journal} {\bibinfo  {journal} {J. Classif.}\ ,\ \bibinfo {pages} {1}}
  (\bibinfo {year} {2019})}\BibitemShut {NoStop}%
\bibitem [{\citenamefont {Hanley}\ and\ \citenamefont
  {McNeil}(1982)}]{Hanley1982Apr}%
  \BibitemOpen
  \bibfield  {author} {\bibinfo {author} {\bibfnamefont {J.~A.}\ \bibnamefont
  {Hanley}}\ and\ \bibinfo {author} {\bibfnamefont {B.~J.}\ \bibnamefont
  {McNeil}},\ }\href {\doibase 10.1148/radiology.143.1.7063747} {\bibfield
  {journal} {\bibinfo  {journal} {Radiology}\ } (\bibinfo {year} {1982}),\
  10.1148/radiology.143.1.7063747}\BibitemShut {NoStop}%
\bibitem [{\citenamefont {L{\ifmmode\ddot{u}\else\"{u}\fi}}\ and\ \citenamefont
  {Zhou}(2011)}]{Lu2011Mar}%
  \BibitemOpen
  \bibfield  {author} {\bibinfo {author} {\bibfnamefont {L.}~\bibnamefont
  {L{\ifmmode\ddot{u}\else\"{u}\fi}}}\ and\ \bibinfo {author} {\bibfnamefont
  {T.}~\bibnamefont {Zhou}},\ }\href {\doibase 10.1016/j.physa.2010.11.027}
  {\bibfield  {journal} {\bibinfo  {journal} {Physica A}\ }\textbf {\bibinfo
  {volume} {390}},\ \bibinfo {pages} {1150} (\bibinfo {year}
  {2011})}\BibitemShut {NoStop}%
\end{thebibliography}

\clearpage

\onecolumngrid
\begin{center}

    \textsc{\Large{Supplementary Information}}
\end{center}

\section{Data collection}
\label{SM:data}

The overlap between APS and ArnetMiner databases leads to a data set of 248,249 papers, authored by 172,589 authors and with 16,175 different keywords.

\subsection{Network reconstruction}

\begin{figure}[tbp]
\centering
\includegraphics[width=0.8\columnwidth]{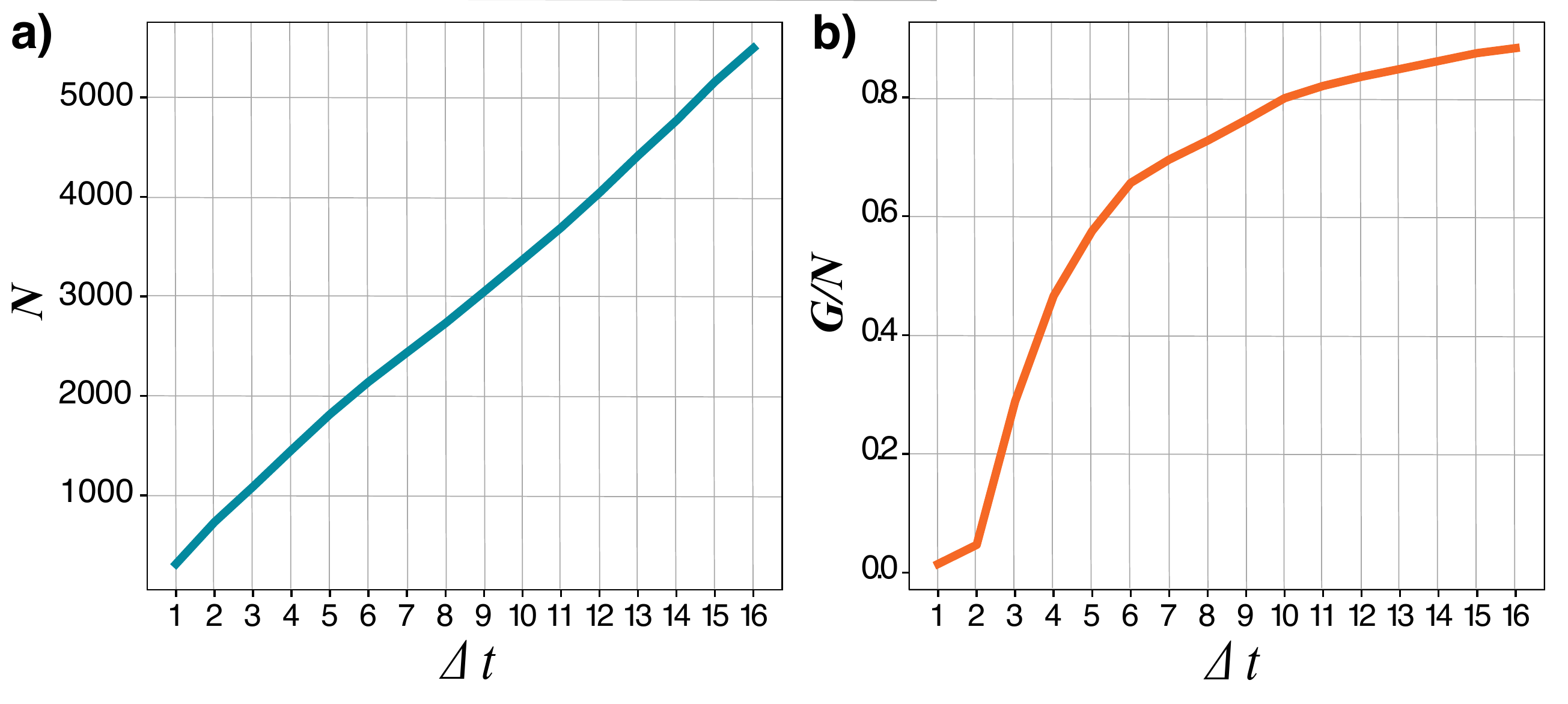}\\
\caption{\textbf{(a)} Number of nodes $N$ of the collaboration network with respect of the time span width $\Delta t$ during the time period 1994-2009. \textbf{(b)} Variation in the fraction of nodes belonging to the giant connected component. Increasing the width of the years range considered, the giant component grows quite fast, reaching almost the whole network.}
\label{fig:N_GN}
\end{figure}

First, we limit the maximum number of authors. Scientific collaboration networks represent social interactions and not a merely co-presence of authors in a paper. Hence, we set a limit for the number of authors in the papers we want to consider: publications with more than 10 authors are excluded. It is reasonable to think that in working groups that are bigger, people are less likely to have real social interactions. Secondly, we need to select the time window for reconstructing the network. As starting year, we choose 1994, which is the first year in which \textit{Physical Review E} is published along other APS journals. Then, in order to select a proper time range, we look at the network's connectedness with respect to the width of the time window, $\Delta t$.
Fig. \ref{fig:N_GN} shows the network's size $N$ and its giant connected component $G$ as a function of $\Delta t$, from 1 year (only 1994) to 16 years (from 1994 to 2009 included). Fig. \ref{fig:N_GN} shows that while $N$ increases linearly with $\Delta t$, the giant connected component increases much faster with $\Delta t$ and then saturates, as usual in percolation processes. Therefore, we selected a time window of 6 years, from 1994 to 2000, corresponding to the point where $G(\Delta t)$ is larger than 50\% of the network size and the giant component starts to saturate. Finally, in order to reduce the number of nodes and thus the computational complexity of link prediction algorithms, we select only papers published within \textit{Physical Review Letters} (PRL). 

\subsection{Keywords cleaning}
\label{SM:kws}
In order to have a uniform and coherent set of keywords used in PRL publications, we started by cleaning such a set. The first step was to automatically convert every letter in lowercase and remove the possible white spaces before and/or after the words. Then, for each keyword, we manually checked the presence of different versions. If present, we converted all of them into a chosen one and removed possible duplicates (in some papers, for example, both the singular and the plural versions of the same word were present). In general, we followed four main rules to perform the selection.

\begin{itemize}
  \item  Set to plural all elementary particles names. With few exceptions (e.g. \emph{higgs boson}), we considered only the plural form for all particles.

  \item Uniform singular/plural. A lot of words appeared both in the singular and in the plural form depending on the paper. We decided in almost every case to keep the most used form between the two.
  
  \item Uniform spelling differences. The same word or construction was written with some spelling differences from one paper to another, e.g., \emph{monte carlo} vs. \emph{montecarlo}, \emph{x~ray} vs. \emph{x-ray}. We chose one version and uniformed all to this one.
  
  \item Uniform UK English vs. American English. Spelling differences were present (e.g. \emph{aluminium} vs. \emph{aluminum}). In these cases we usually kept the UK English form.
  
\end{itemize}

\begin{figure}[tbp]
\centering
\includegraphics[width=0.6\textwidth]{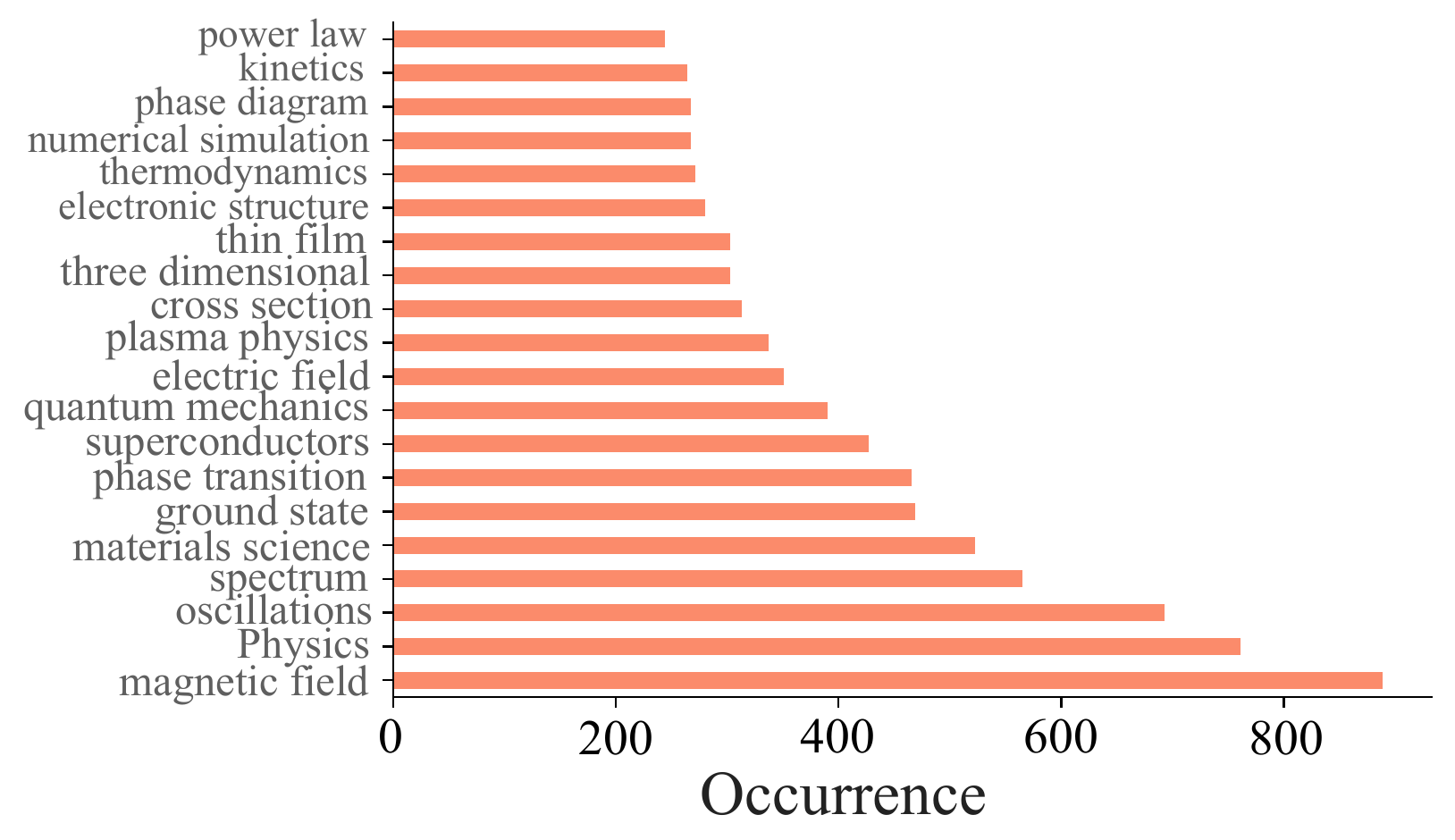}
\caption{\textbf{Top 20 occurring keywords}. The 20 most occurring keywords in PRL papers published during the time window 1994-2000.}
\label{fig:kws}
\end{figure}

At the end of this process, we obtained a set of 7239 different keywords. Figure \ref{fig:kws} shows the occurrence of the top 20 keywords used in the papers selected. The second most common keyword is \emph{physics}, which we excluded from the set. 

\section{The limitations of similarity-based techniques \label{sec:single_layer}}

The quality of the predictions is not only determined by the metrics used to compute the score, but also by the availability of information on its own, as pointed out by Jia et al. \cite{Jia2020}. For instance, in sparse networks, such as most real-world networks, most pairs of nodes will be without common neighbors and will be assigned exactly the same score, zero. In general, there will always be a set of scoreless links, limiting the maximum and minimum values of the AUC measure, to
\begin{equation}\label{eq:AUC_worst}
\text{AUC}_\text{min} = \frac{1}{2}(1+p_1)(1-p_2)  \qquad  \text{AUC}_\text{max} = \text{AUC}_\text{min} + p_1 p_2\,,
\end{equation}
where $p_1$ is the fraction of links with a score different from 0 among those links that will exist in the future, and $p_2$ is the same among the edges that will not exist.
Only in case of no scoreless links, that is when $p_1 = p_2 = 1$, we obtain $\text{AUC}_\text{min} = 0$ and $\text{AUC}_\text{max} =1$.

The above equation can be derived as follows. The limited amount of links present in a network bounds the prediction power of any similarity-based link prediction algorithm. This is due to the fact that scoreless links limit the value of AUC. The problem of having equal scores is not trivial and different statistical software packages choose different ways of solving the ties \cite{Muschelli2019Dec}. If we consider the classical approach of solving them (i.e., all links with the same score produce a single point in the TPR/FPR curve), the AUC is equivalent to the Mann Whitney Wilcoxon test \cite{Hanley1982Apr},
 so it reads:
\begin{equation}\label{eq:AUC_wilcoxon}
\text{AUC} = \frac{n'+0.5n''}{n}\,.
\end{equation}
The meaning of this expression is as follows. First, we pick two random links, one from the set of links that will exist in the future, which we denote as $P_1$, and one from the set of links that will not exist in the future, $P_2$. 
If the score of the link belonging to the first group is larger than the one from the second, $n'$ is incremented by 1.
If the score is the same, $n''$ is incremented instead. If this process is repeated $n$ times, equation \eqref{eq:AUC_wilcoxon} is equivalent to the classical area under the curve measured from the TPR/FPR curve \cite{Lu2011Mar}.

Following \cite{Jia2020}, the bounds on the AUC can be obtained by measuring the fraction of links in the set $P_1$ with a score different from 0, $p_1$, and similarly for the set $P_2$, yielding $p_2$. Hence, the fraction of links with a score equal to 0 in both sets will be $(1-p_1)$ and $(1-p_2)$, respectively. Thus, in the worst case scenario in which all links corresponding to $p_2$ have a score larger than $p_1$ we would have $n'/n=p_1(1-p_2)$ and $n''/n=(1-p_1)(1-p_2)$, so that
\begin{equation}\label{eq:AUC_worst}
\text{AUC}_\text{worst} = \frac{1}{2}(1+p_1)(1-p_2)\,.
\end{equation}
A similar argument allows us to determine the best possible AUC, i.e., the one in which all the links corresponding to $p_1$ have a score larger than the ones from $p_2$ (note that there will be a fraction $1-p_1$ of links that will exist in the future with score equal to 0 and hence below the ones corresponding to $p_2$) yielding
\begin{equation}\label{eq:AUC_best}
\text{AUC}_\text{best} = \text{AUC}_\text{worst} + p_1 p_2\,.
\end{equation}

\section{Link prediction metrics}

To determine if the information provided by the citation and keyword layers is actually useful to predict the appearance of new links in the collaboration layer, we propose two novel metrics based on the similarity between nodes in these layers.
\begin{itemize}
\item Mutual citations ($MC$): if two authors mutually cite each other, it might be more likely for them to collaborate. 
This MC score between nodes $u$ and $v$ is defined simply as the weight of the link between $u$ and $v$ in the citation layer,
\begin{equation}\label{eq:MC}
MC(u,v) = w^r_{uv}\,.
\end{equation}

\item Common keywords ($CK$): similarly, if two authors show common scientific interests, using the same set of keywords, the chances that they collaborate in the future should be higher than if they did not have common interests. Thus, the $CK$ score between nodes $u$ and $v$  can be expressed as the weight of a link between $u$ and $v$  in the keyword layer,
\begin{equation}\label{eq:CK}
CK(u,v) = w^{k}_{uv}\,.
\end{equation}
\end{itemize}

For each case, we also try a normalized variant:
\begin{itemize}
\item Normalized mutual citations ($NMC$): the number of references between two authors should be related to the total credit received by each of them. Then, the score becomes
\begin{equation}\label{eq:NMC}
NMC(u,v) = \frac{w^r_{uv}}{s^r_u}+\frac{w^r_{vu}}{s^r_v}\,.
\end{equation}
The normalization terms $s^r_u$ and $s^r_v$ are the total number of citations received by $u$ and $v$, respectively (i.e., the total incoming strength of the two nodes). Indeed, to apply this method we consider the directed behavior of the citations network, explicitly differentiating for each author the received citations and the given references. 
\item Normalized common keywords ($NCK$): some authors have a larger list of keywords than others, so they are more likely to share keywords with someone else. To verify if this is a relevant element for the prediction performance, the score is computed as
\begin{equation}\label{eq:NCK}
NCK(u,v) = \frac{w^k_{uv}}{\text{max}(K_u,K_v)}\,,
\end{equation}
where $K_u$ is the whole list of keywords used by $u$ and $\text{max}(K_u,K_v)$ represents the maximum between $K_u$ and $K_v$.
\end{itemize}

\begin{figure}
\centering
\includegraphics[width=0.6\textwidth]{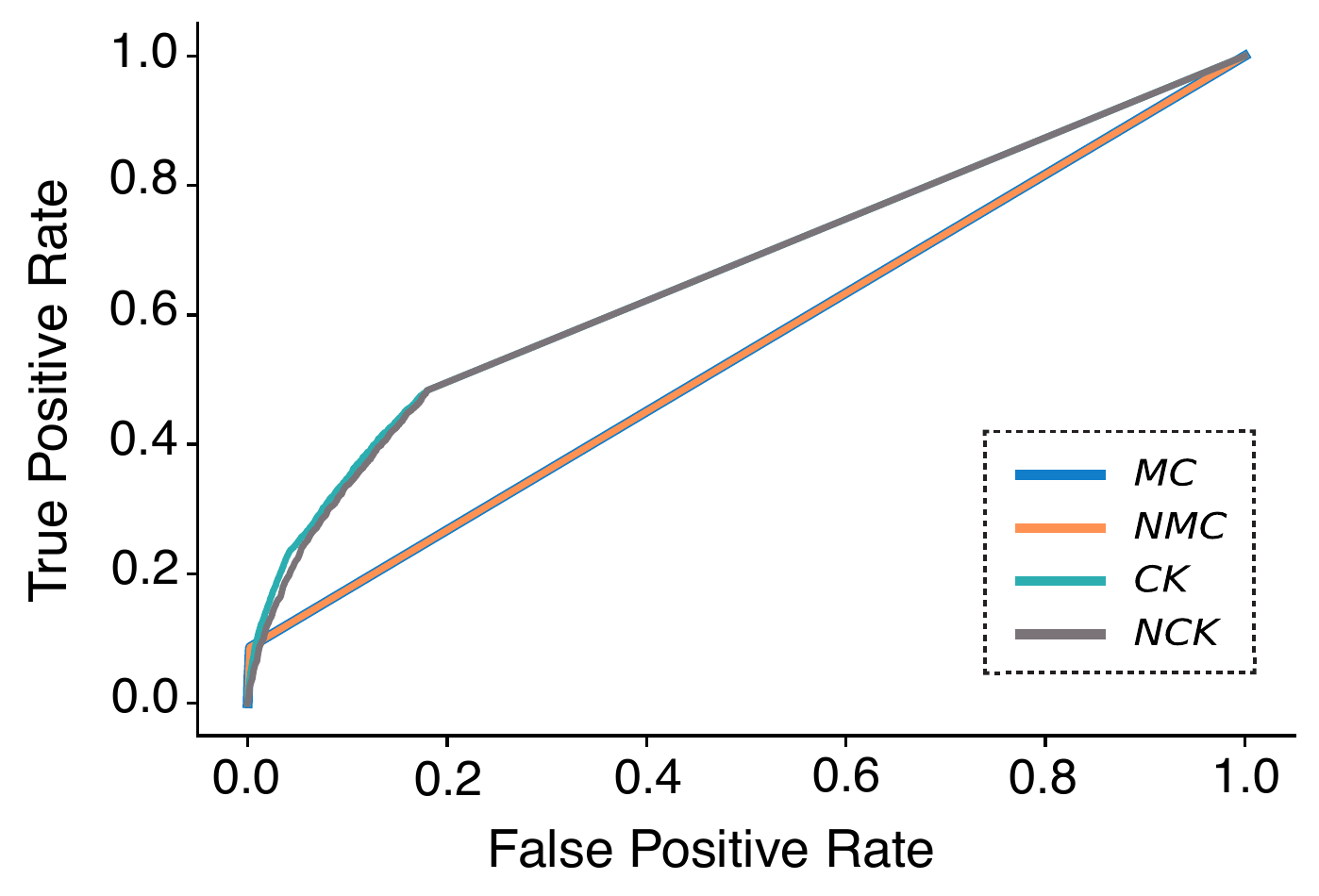}
\caption{\textbf{Comparison of the ROC curves corresponding to the two variants of each method}. The almost complete overlap between the curves shows that the different versions of the metrics are not really affecting the prediction.}
\label{fig:variants}
\end{figure}

\begin{table}[]
\centering
\begin{tabular}{lccc}
\hline
\textbf{Method}	& \textbf{Precision} & \textbf{AUC} 	& \textbf{AUC [worst-best]} \\
\hline 
$MC$                 	& 0.025 		& 0.5421           & [0.5420-0.5422]  \\
$NMC$ 			& 0.023 		& 0.5422           & [0.5420-0.5422] \\
$CK$             		& 0.012 		& 0.6648           & [0.6082-0.6952] \\
$NCK$                     	& 0.005 		& 0.6618           & [0.6082-0.6952] \\
\hline \\
\end{tabular}
\caption{Precision values, AUC values and theoretical bounds of the AUC obtained for each of the proposed metrics. Both the normalized versions give almost the same outcome than the original versions.}
\label{tab:variants}
\end{table}

Figure \ref{fig:variants} and Table \ref{tab:variants} show the results of the application of these methods.
We can observe that there is a very small difference between the two versions of each method, particularly for the citations network. This outcome enables us to confirm that the use of an undirected network to depict references does not affect the prediction performance. 
Note that the bounds of the AUC value are the same whether the methods are normalized or not. This aspect is due to the fact that such limitations are determined by the amount of information provided by the network and do not depend on the specific shape of the metric.
Compared with the outcome obtained with the Adamic-Adar method applied on citations layer, $MC$ and $NMC$ have higher Precision but smaller AUC value. With $CK$ and $NCK$ we obtain higher AUC values than $AA_k$ and higher Precision in one case, while slightly smaller in the other one.
However, none of these methods can outperform the Precision of the $AA_c$ score, even if $CK$ and $NCK$ can reach higher AUC values (see main text).

\section{Adamic-Adar score for multiplex network}

\begin{figure}
\centering
\includegraphics[width=0.6\textwidth]{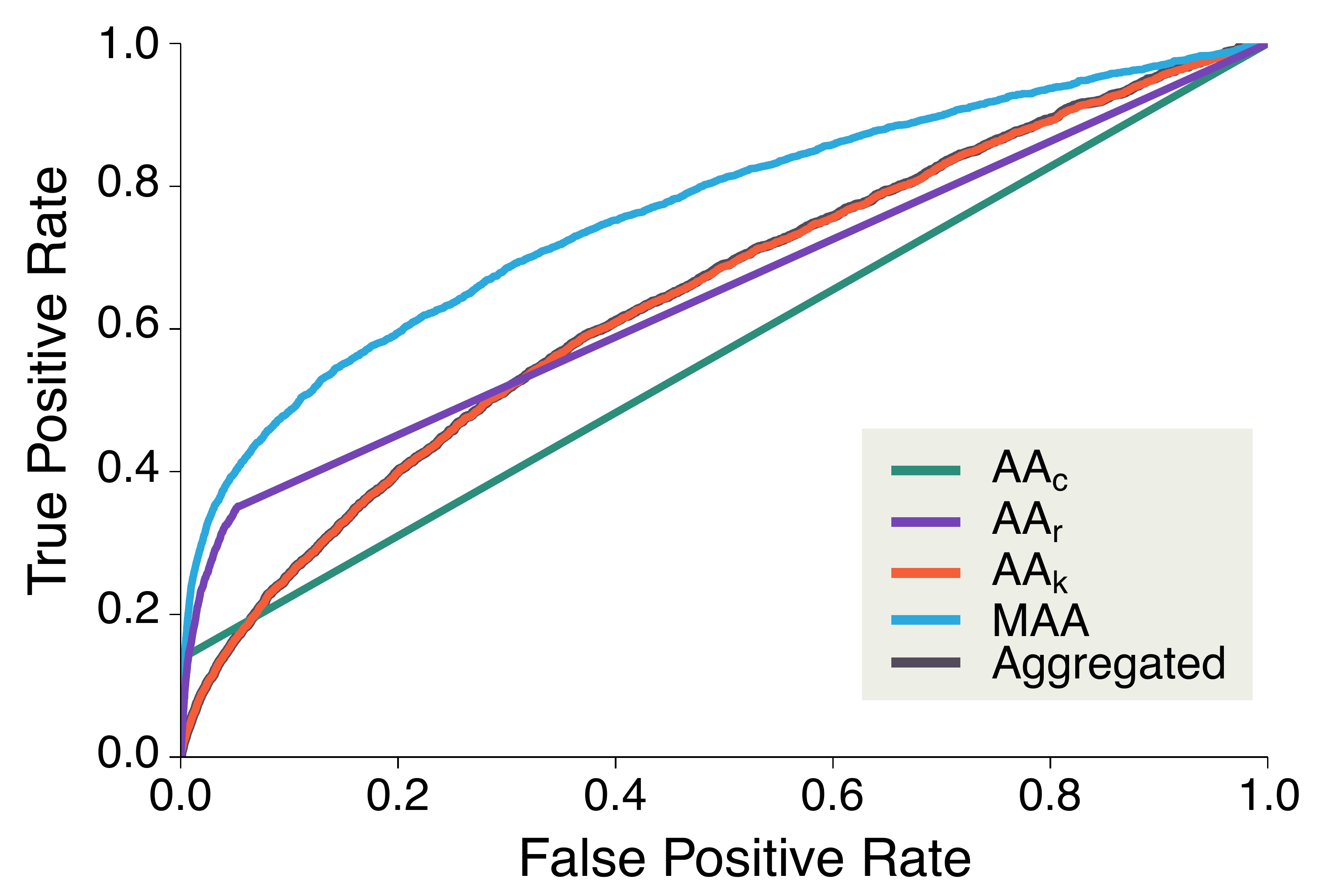}
\caption{\textbf{ROC curves obtained for different types of triadic closure, for $MAA$ and for the aggregated network when $k_{min}=5$.}
In general, the behavior of the curves is almost the same of the case with a larger \emph{Core}.}
\label{fig:ROC_multiplex_5}
\end{figure}

In order to verify whether or not the number of nodes and new links to predict can influence the performance, we set another value for $k_{min}$ and repeat the same simulations shown in the main text. In particular, we consider $k_{min}=5$ obtaining $E_p=4491$ new collaborations to predict, among 3599 authors in the \emph{Core}. Figure \ref{fig:ROC_multiplex_5} shows the ROC curves obtained from the application of the $AA$ method to the multiplex structure. Table \ref{tab:AUC_multiplex_5} shows the numerical results. Although the values and the shapes of the curves slightly change as compared to the case of $k_{min}=3$, in general, we see that the performances of the metrics are not really affected by the dimension of the \emph{Core}. 

\begin{table}[]
\centering
\begin{tabular}{lccc}
\hline
\textbf{Configuration}	& \textbf{Precision} & \textbf{AUC} & \textbf{AUC [worst-best]} \\
\hline
$AA_c$   	& 0.048                   & 0.5686        & [0.5682-0.5688]                 \\
$AA_r$    & 0.020                   & 0.6527        & [0.6391-0.6582]                 \\
$AA_k$   & 0.008                    & 0.6433        & [0.0106-0.9969]               \\
$MAA$ (all triads)      & 0.046                    & 0.7634        & [0.0079-0.9984]              \\
Aggregated               & 0.008                     & 0.6443        & [0.0079-0.9984]              \\
\hline \\
\end{tabular}
\caption{Precision and AUC values obtained for different types of triangles, as well as the theoretical bounds of the AUC, with $k_{min}=5$.
In $AA_c$, $AA_r$ and $AA_k$ each layer is considered separately. The $MAA$ score is computed using parameters $\eta_{ck}=0.05$ and $\eta_{cr}=0.15$.}
\label{tab:AUC_multiplex_5}
\end{table}

\begin{figure}
\centering
\includegraphics[width=\textwidth]{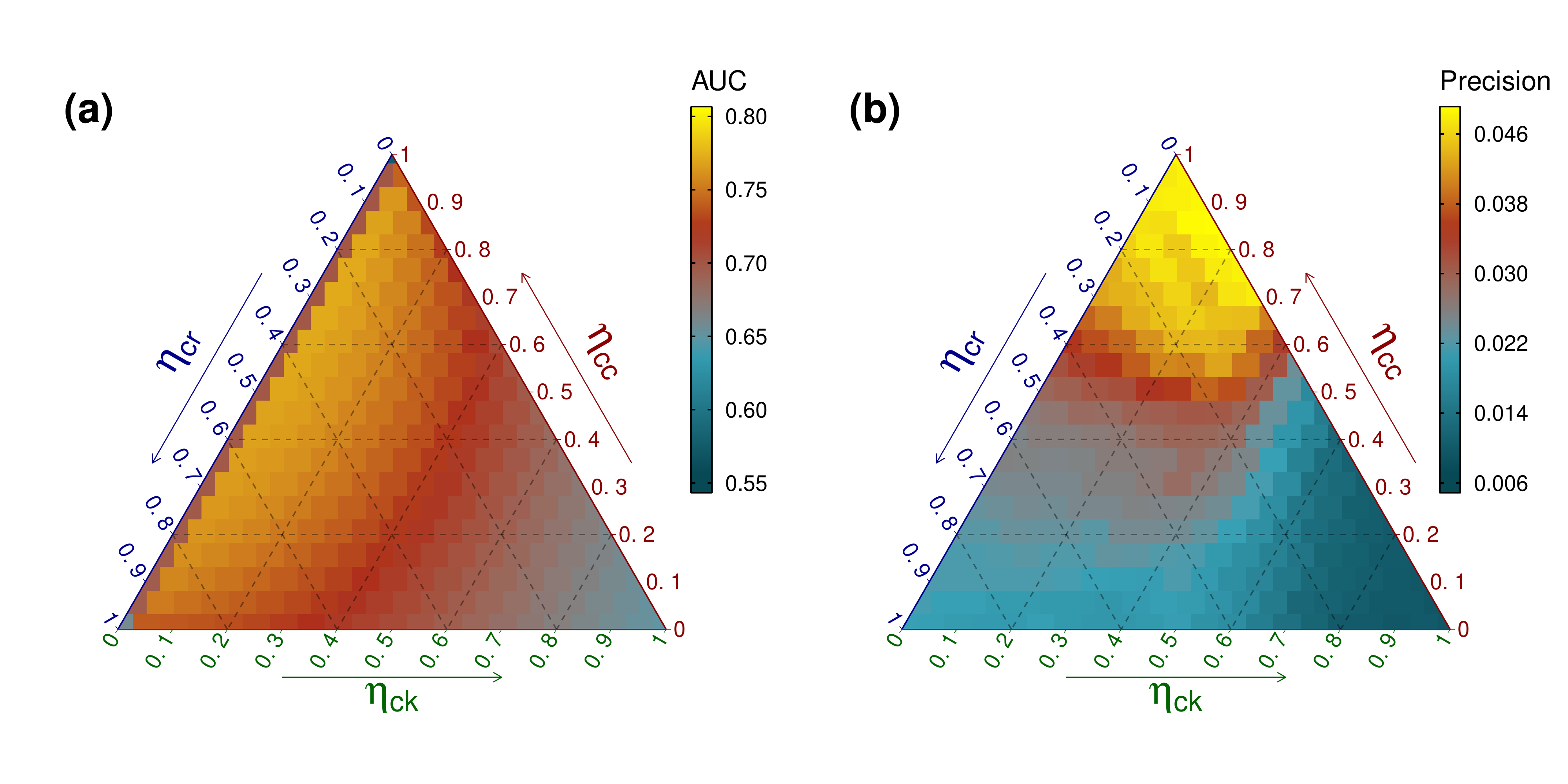}
\caption{\textbf{AUC and Precision trends obtained when $k_{min}=5$.} We vary the values of the parameters $\eta_{cr}$ and $\eta_{ck}$, which consequently, fixes the third parameter $\eta_{cc}$. Panel (a) shows that AUC values grow fast when we introduce types of triangles involving also layer $r$ and layer $k$, and then decrease slowly. Panel (b) depicts the importance of the $\mathcal{T}_{cc}$ triads for the Precision, which is high for small values of $\eta_{cr}$ and $\eta_{ck}$ and then decreases. Both AUC and Precision have a similar behavior to the one reported for $k_{min}=3$ in the main text.}
\label{fig:phsp_5}
\end{figure}

Figure \ref{fig:phsp_5} depicts the trend of the prediction performance obtained using $MAA$ with different values of the parameters.
The behavior is similar to the one obtained when $k_{min}=3$. As before, we can see a discontinuity in AUC values when the contribution of triangles different from $AA_c$ is introduced, due to the addition of further information. In general, both for AUC and Precision, the values obtained in this case are slightly higher than the ones obtained with more nodes in the \emph{Core}. This may be related to the fact that having $k\ge 5$ probably means that the author is more active, and hence also the citations and keywords networks can provide more information.

\end{document}